\documentclass[onecolumn,epjc3]{svjour3}
\usepackage{amsfonts}
\usepackage{mathrsfs}
\usepackage{dblfnote}
\usepackage{graphicx}
\usepackage{hyperref}
\usepackage{bm}
\usepackage{amsmath}
\DeclareMathAlphabet{\mathscrbf}{OMS}{mdugm}{b}{n}
\usepackage{geometry}
\usepackage{indentfirst}
\usepackage{color}
\usepackage{acro}
\usepackage{lineno}
\usepackage{cancel}
\usepackage[numbers,sort&compress]{natbib}

\DeclareAcronym{CMB}{
    short = CMB,
    long = Cosmic Microwave Background Radiation,
}

\DeclareAcronym{PTA}{
    short = PTA,
    long = pulsar timing arrays,
}

\DeclareAcronym{SKA}{
    short = SKA,
    long = the Square Kilometer Array,
}

\begin{document}


\bibliographystyle{unsrt}

\title{Impact of the free-streaming neutrinos to the second order induced gravitational waves}

\author{
  Xukun Zhang\thanksref{e1,addr1,addr2} \and
  Jing-Zhi Zhou\thanksref{e2,addr1,addr2} \and
  Zhe Chang\thanksref{e3,addr1,addr2}
}

\thankstext{e1}{e-mail: zhangxukun@ihep.ac.cn}
\thankstext{e2}{e-mail: zhoujingzhi@ihep.ac.cn}
\thankstext{e3}{e-mail: changz@ihep.ac.cn}

\institute{Institute of High Energy Physics, Chinese Academy of Sciences, 100049 Beijing, China\label{addr1}\and 
University of Chinese Academy of Sciences, Beijing 100049, China\label{addr2}
}
\date{Received: date / Accepted: date}

\maketitle

\setlength{\parindent}{1.7em}

\begin{abstract}
    The damping effect of the free-streaming neutrinos on the second order gravitational waves is investigated in detail. We solve the Boltzmann equation and give the anisotropic stress induced by neutrinos to second order. The first order tensor and its coupling with scalar perturbations induced gravitational waves are considered. We give the analytic equations of the damping kernel functions and finally obtain the energy density spectrum. The results show that the free-streaming neutrinos suppress the density spectrum significantly for low frequency gravitational waves and enlarge the logarithmic slope $n$ in the infrared region ($k \ll k_*$) of the spectrum. For the spectrum of $k_*\sim 10^{-7}$Hz, the damping effect in the range of $k<k_*$ is significant. The combined effect of the first and second order could reduce the amplitude by $30\%$ and make $n$ jump from $1.54$ to $1.63$ at $k\sim 10^{-9}$Hz, which may be probed by the pulsar timing arrays (PTA) in the future. 
  \keywords{neutrino \and induced gravitational waves}
\end{abstract}

\section{Introduction}\label{sec:intro}


Since the detection of gravitational waves by the LIGO and Virgo collaborations \cite{LIGOScientific:2016aoc, LIGOScientific:2016sjg}, gravitational waves have been a popular research object of the cosmology. In particular, the induced gravitational waves have attracted great attention. For recent review please see Ref.~\cite{Domenech:2021ztg} and references therein. On large scales, \ac{CMB} gives us information of the primordial fluctuations and inflation \cite{Planck:2018vyg, Planck:2018jri, Cai:2019bmk}. In contrast, the induced gravitational waves, generated from the primordial quantum fluctuations, could be used to study the early Universe and new physics beyond the Standard Model on small scales. It contains the information about the physics of the early universe, such as the primordial black holes \cite{Saito:2008jc, Wang:2016ana, Wang:2019kaf, Yuan:2019udt} and the primordial non-Gaussianity \cite{Garcia-Bellido:2016dkw, Cai:2018dig}. The induced gravitational waves from the matter dominated \cite{Inomata:2019zqy, Inomata:2019ivs, Assadullahi:2009nf} and the primordial black hole dominated era \cite{Papanikolaou:2020qtd, Domenech:2020ssp} have been studied. More recently, the gauge issue \cite{Chang:2020mky, Chang:2020iji, Chang:2020tji, Hwang:2017oxa} and the higher order effects \cite{Yuan:2019udt,Zhou:2021vcw, Chang:2022dhh} were analyzed. 

It is well known that the neutrinos and the other cosmic particles could have significant impacts on the evolution of the primordial gravitational waves. Considering the anisotropic stress from the free-streaming neutrinos, we know that the square amplitude of the primordial gravitational waves could be reduced by about $35.6\%$ \cite{Weinberg:2003ur,Dent:2013asa, Watanabe:2006qe}. For the second order scalar induced gravitational waves, the effects of the free-streaming neutrinos have been studied in Ref.~\cite{Bartolo:2010qu, Mangilli:2008bw, Saga:2014jca}. At the same time, there are studies \cite{Nakama:2016enz, Gong:2019mui} concerning the tensor induced perturbations independent of neutrinos.
In this paper, the first order scalar and tensor perturbations are considered. We give the kernel functions contributed by the scalar-scalar, scalar-tensor and tensor-tensor coupled source terms with impacts of neutrinos for the first time. Besides, we take account of the omitted terms like the integrates of $\gamma^{i_1}...\gamma^{i_n}\phi^{(1)}\psi^{(1)}, (n\neq 0, 2)$ in the second order anisotropic stress $\Pi_{ij}^{(2)}$ \cite{Mangilli:2008bw}. 
Finally, giving a monochromatic primordial power spectrum, we obtain the density spectrum of the gravitational waves induced by scalar.

The remaining part of this paper is organized as follows. In Sec.~\ref{sec:cal}, we solve the Boltzmann equation and obtain the anisotropic stress up to the second order induced by neutrinos. In Sec.~\ref{sec:trans_and_kernel} we give the equation of motion of second order gravitational waves and obtain the kernel functions. The solutions of the equations are shown and analyzed. Finally, in Sec.~\ref{sec:energy_spectrum}, we calculate and show a comparison of energy density spectrum between no damping and damping gravitational waves. Conclusion and discussions are presented in Sec.~\ref{sec:con}.

\section{The anisotropic stress induced by neutrinos}\label{sec:cal}
The perturbed Friedmann-Robertson-Walker (FRW) metric up to second order in Newtonian gauge is given by
\begin{equation}\label{eq:metric_t}
    {\rm d}s^2 = -(1+2\phi^{(1)}+\phi^{(2)}){\rm d}t^2 + a(t)V^{(2)}e_i {\rm d}t{\rm d}x^i+a^2(t)\left((1-2\psi^{(1)}-\psi^{(2)})\delta_{ij}+h_{ij}^{(1)}+\frac{1}{2}h_{ij}^{(2)}\right){\rm d}x^i{\rm d}x^j \ ,
\end{equation}
where $\partial^i e_i = \partial^i h_{ij}^{(n)} = \delta^{ij}h_{ij}^{(n)} = 0 \ (n=1,2)$. In eq.~(\ref{eq:metric_t}), we have neglected the first order vector perturbation for the reason that it has a decreasing amplitude and can not be generated in the presence of scalar field \cite{Bartolo:2004if,Mangilli:2008bw}.

Ref.~\cite{Dent:2013asa} has taken into account neutrino masses for $\mathcal{O}(1)$ eV and found that the influence is very small (see Fig.~1 in Ref.~\cite{Dent:2013asa}). Therefore, in this paper, the neutrino mass has been neglected. The four momentum of neutrino is defined as
\begin{equation}
    P^\mu \equiv \frac{{\rm d}x^\mu}{{\rm d}\lambda} \ ,
\end{equation}
where $\lambda$ is an affine parameter. And $P^\mu$ satisfy the constraint $g_{\mu\nu}P^\mu P^\nu=0$. Then we could define the three momentum ${\bm p}$ as
\begin{align}
    p^2 &= g_{ij}P^iP^j \ ,\label{eq:p^2} \\
    p^i &= \gamma^i p \ \label{eq:p^i},
\end{align}
where $p=\sqrt{\delta_{ij}p^ip^j}$ is the module of ${\bm p}$, ${\bm \gamma}$ denotes the direction of ${\bm p}$ and $\delta_{ij}\gamma^i\gamma^j=1$. Here, we have set ${\bm p}$ parallel with ${\bm P}$.

The Boltzmann equation for decoupled neutrinos can be written as
\begin{equation}\label{eq:Boltzmann_equation}
    \frac{{\rm d}F(q, {\bm \gamma}, {\bm x}, \eta)}{{\rm d}\eta} = 0 \ ,
\end{equation}
where $F$ is the distribution function of neutrino, ${\bm q}\equiv a{\bm p}$ is the comoving three momentum and $\eta$ is the conformal time. We present $F$ to second order 
\begin{equation}
    F(q, {\bm \gamma}, {\bm x}, \eta) = F^{(0)}(q) + F^{(1)}(q, {\bm \gamma}, {\bm x}, \eta) + \frac{1}{2}F^{(2)}(q, {\bm \gamma}, {\bm x}, \eta) \ , 
\end{equation}
where $F^{(0)}$ is given by the Fermi-Dirac distribution.

The free-streaming neutrinos would produce an anisotropic stress, namely
\begin{equation}
    \Pi_{ij} = \Pi_{ij}^{(1)} + \frac{1}{2} \Pi_{ij}^{(2)} \ .
\end{equation}
We can decompose the stress by helicity
\begin{equation}\label{eq:decomposition_of_pi}
    \Pi_{ij}^{(n)} = \sigma_{ij}^{\mathrm{TT}(n)}+\frac{1}{2}(\partial_i \sigma_j^{(n)}+\partial_j \sigma_i^{(n)})+\left(\partial_i \partial_j-\frac{1}{3}\delta_{ij}\Delta\right)\sigma^{(n)}\ ,\quad (n=1,2) \ ,
\end{equation}
where $\partial^i\sigma_i^{(n)}=\partial^i \sigma_{ij}^{\mathrm{TT}(n)}=\delta^{ij}\sigma_{ij}^{\mathrm{TT}(n)}=0$.

The anisotropic stress from neutrinos can be expressed in terms of the distribution function \cite{Mangilli:2008bw, Watanabe:2006qe}
\begin{equation}\label{eq:Pi_ij_n}
    \Pi_{ij}^{(n)} = a^{-4}\int \frac{{\rm d}^3 q}{(2\pi)^3} q\gamma_i\gamma_j F^{(n)}\ ,\quad (n=1,2) \ .
\end{equation}

\subsection{The first-order Boltzmann equation}\label{cal:1st}
In \ref{sec:geodesic}, we showed that the zero-order distribution function of neutrinos satisfies the identity $(\frac{\partial F}{\partial x^i})^{(0)}=(\frac{{\rm d} q}{{\rm d} \eta})^{(0)}=(\frac{{\rm d} \gamma^i}{{\rm d} \eta})^{(0)}=(\frac{\partial F}{\partial \gamma^i})^{(0)}=0$. Therefore, the first-order Boltzmann equation can be written as
\begin{equation}\label{eq:1st_be}
    \left(\frac{\partial F}{\partial \eta}\right)^{(1)} + \left(\frac{{\rm d} x^i}{{\rm d} \eta}\right)^{(0)}\left(\frac{\partial F}{\partial x^i}\right)^{(1)} + \left(\frac{{\rm d} q}{{\rm d} \eta}\right)^{(1)}\left(\frac{\partial F}{\partial q}\right)^{(0)} = 0 \ .
\end{equation}
Using Fourier transform, we rearrange the eq.~(\ref{eq:1st_be}) in the form of
\begin{equation}
    \frac{\partial f^{(1)}_{\bm k}}{\partial \eta} + i\gamma^i k_i f^{(1)}_{\bm k} + q(-i\gamma^i k_i \phi_{\bm k}^{(1)}+\partial_\eta \psi_{\bm k}^{(1)}-\frac{1}{2}\gamma^i\gamma^j\partial_\eta h_{{\bm k},ij}^{(1)})\left(\frac{\partial F}{\partial q}\right)^{(0)} = 0 \ ,
\end{equation}
where $f_{\bm k}^{(1)}$, $\phi_{\bm k}^{(1)}$, $\psi_{\bm k}^{(1)}$, and $h_{{\bm k},ij}^{(1)}$ denote the counterparts of $F^{(1)}$, $\phi^{(1)}$, $\psi^{(1)}$, and $h_{ij}^{(1)}$ in momentum space, respectively. In this paper, the Fourier transforms of the quantities are introduced as 
\begin{equation}
    X^{(n)}(\eta,{\bm x}) = \int \frac{d^3 k}{(2\pi)^{3/2}} X_{\bm k}^{(n)}(\eta) e^{i{\bm k}{\cdot}{\bm x}} \ ,
\end{equation}
where $X = F$, $\phi$, $\psi$, $h_{ij}$, $\sigma$ ... and $n = 1, 2$.

The first order Boltzmann equation has been studied in the previous work \cite{Mangilli:2008bw, Weinberg:2003ur, Watanabe:2006qe}. The solution is given by
\begin{equation}
    f_{\bm k}^{(1)} = q\left(\frac{\partial F}{\partial q}\right)^{(0)}\left(\phi_{\bm k}^{(1)}(\eta) + \int_{\eta_\mathrm{dec}}^\eta {\rm d}\eta'\left(\frac{1}{2}\gamma^i\gamma^j\epsilon_{{\bm k},ij}^\lambda \partial_\eta h_{{\bm k},\lambda}^{(1)}(\eta') -\partial_\eta\phi_{\bm k}^{(1)}(\eta') - \partial_\eta\psi_{\bm k}^{(1)}(\eta') \right)e^{-i\gamma^l k_l(\eta-\eta')} \right) \ ,
\end{equation}
where $\eta_\mathrm{dec}$ is the time of neutrino decoupling, $\epsilon_{{\bm k},ij}^\lambda$ $(\lambda = +,\times)$ are the Fourier modes of the polarization tensors $\epsilon_{ij}^\lambda$ \cite{book:2362664}.
Following the decomposition of the anisotropic stress $\Pi_{ij}$ in eq.~(\ref{eq:decomposition_of_pi}), we obtain the explicit expressions of $\sigma^{(1)}$ and $\sigma_\lambda^{(1)} = \epsilon_{\lambda}^{ij}\sigma_{ij}^{\mathrm{TT}(1)}$ in momentum space,
\begin{align}
    \sigma_{\bm k}^{(1)} &= \frac{1}{k^2}\rho_\nu^{(0)}(\eta)\int_{\eta_\mathrm{dec}}^\eta {\rm d}\eta^\prime \left[j_0(k(\eta-\eta'))-\frac{3j_1(k(\eta-\eta'))}{k(\eta-\eta')}+3j_2(k(\eta-\eta'))\right] \left( \partial_\eta \phi_{\bm k}^{(1)}(\eta^\prime) + \partial_\eta \psi_{\bm k}^{(1)}(\eta^\prime)\right) \ , \label{eq:sigma1_s} \\
    \sigma_{\lambda,\bm k}^{(1)} &= -4\rho_\nu^{(0)}(\eta) \int_{\eta_\mathrm{dec}}^\eta {\rm d}\eta^\prime \frac{j_2(k(\eta-\eta^\prime))}{(k(\eta-\eta^\prime))^2} \partial_\eta h_{\lambda,\bm k}^{(1)}(\eta^\prime) \ , \label{eq:sigma1_t} 
\end{align}
where $\rho_\nu^{(0)}$ is the unperturbed neutrino energy density, $j_n(x)$ is the $n$-order spherical Bessel function.

\subsection{The second-order Boltzmann equation}\label{cal:2nd}
The second-order Boltzmann equation is given by
\begin{equation}
    \begin{split}
        \left(\frac{\partial F}{\partial \eta}\right)^{(2)} + \left(\frac{{\rm d} x^i}{{\rm d} \eta}\right)^{(0)}\left(\frac{\partial F}{\partial x^i}\right)^{(2)} + &\left(\frac{{\rm d} x^i}{{\rm d} \eta}\right)^{(1)}\left(\frac{\partial F}{\partial x^i}\right)^{(1)} + \left(\frac{{\rm d} q}{{\rm d} \eta}\right)^{(1)}\left(\frac{\partial F}{\partial q}\right)^{(1)}\\ 
        &+ \left(\frac{{\rm d} q}{{\rm d} \eta}\right)^{(2)}\left(\frac{\partial F}{\partial q}\right)^{(0)} + \left(\frac{{\rm d} \gamma^i}{{\rm d} \eta}\right)^{(1)}\left(\frac{\partial F}{\partial \gamma^i}\right)^{(1)} = 0 \ .
    \end{split}
\end{equation}
In momentum space, the equation could be written as
\begin{equation}
    \frac{1}{2}\frac{\partial f_{\bm k}^{(2)}}{\partial \eta} + \frac{1}{2}i\gamma^j k_j f_{\bm k}^{(2)} = A(q,{\bm k},{\bm \gamma},\eta) \ ,
\end{equation}
where $A(q,{\bm k},{\bm \gamma},\eta)$ is defined as
\begin{equation}\label{eq:definition_of_A}
    A(q,{\bm k},{\bm \gamma},\eta) \equiv -\left[ \left(\frac{{\rm d} x^i}{{\rm d} \eta}\right)^{(1)}\left(\frac{\partial F}{\partial x^i}\right)^{(1)} + \left(\frac{{\rm d} q}{{\rm d} \eta}\right)^{(1)}\left(\frac{\partial F}{\partial q}\right)^{(1)} + \left(\frac{{\rm d} \gamma^i}{{\rm d} \eta}\right)^{(1)}\left(\frac{\partial F}{\partial \gamma^i}\right)^{(1)} \right]_{\bm k} - \left(\frac{{\rm d} q}{{\rm d} \eta}\right)^{(2)}_{\bm k}\left(\frac{\partial F}{\partial q}\right)^{(0)} \ .
\end{equation}
Then we obtain
\begin{equation}\label{eq:2o_f}
    f_{\bm k}^{(2)} = 2\int_{\eta_\mathrm{dec}}^\eta {\rm d}\eta^\prime A(q,{\bm k},{\bm \gamma},\eta^\prime) e^{-i\gamma^j k_j(\eta-\eta^\prime)} \ .
\end{equation}
Finally, the second order anisotropic stress of neutrinos can be written in the form 
\begin{equation}\label{eq:2o_sigma1}
    \begin{split}
        \sigma_{\lambda, {\bm k}}^{(2)} = 8\rho_\nu^{(0)} \bigg[\int\frac{{\rm d}^3 k_1 {\rm d}^3 k_2}{(2\pi)^3}&\delta({\bm k}_1+{\bm k}_2-{\bm k})\big[\tilde{D}_{1,\lambda}({\bm k}, {\bm k}_1, {\bm k}_2, \eta)-4\tilde{D}_{2,\lambda}({\bm k}, {\bm k}_1, {\bm k}_2, \eta)+\tilde{D}_{3,\lambda}({\bm k}, {\bm k}_1, {\bm k}_2, \eta)\\
        &+\tilde{D}_{4{\rm \uppercase\expandafter{\romannumeral1}},\lambda}({\bm k}, {\bm k}_1, {\bm k}_2, \eta)\big]+\tilde{D}_{4{\rm \uppercase\expandafter{\romannumeral2}},\lambda}({\bm k},\eta)\bigg] \ .
    \end{split}
\end{equation}
The functions $\tilde{D}_{i,\lambda}\ (i = 1, 2, 3)$ come from the integrals of the terms $\left[\left(\frac{{\rm d} x^i}{{\rm d} \eta}\right)^{(1)} \left(\frac{\partial F}{\partial x^i}\right)^{(1)}\right]_{\bm k}$,$\left[\left(\frac{{\rm d} q}{{\rm d} \eta}\right)^{(1)}\left(\frac{\partial F}{\partial q}\right)^{(1)}\right]_{\bm k}$, $\left[\left(\frac{{\rm d} \gamma^i}{{\rm d} \eta}\right)^{(1)}\left(\frac{\partial F}{\partial \gamma^i}\right)^{(1)}\right]_{\bm k}$, respectively. $\tilde{D}_{4{\rm \uppercase\expandafter{\romannumeral1}},\lambda}$ and $\tilde{D}_{4{\rm \uppercase\expandafter{\romannumeral2}},\lambda}$ are from the integral of $\left(\frac{{\rm d} q}{{\rm d} \eta}\right)^{(2)}_{\bm k}\left(\frac{\partial F}{\partial q}\right)^{(0)}$.
The explicit forms of functions $\tilde{D}_{n,\lambda}$ are defined in eq.~(\ref{eq:def_of_Dtilde}).

Ref.~\cite{Kasai:1985jg} has given a decomposition of the first order distribution function in Fourier space, 
\begin{equation}
    f_{\bm k} = f_{\bm k}^S + \gamma^i e_i f_{\bm k}^V + \gamma^i \gamma^j \epsilon_{{\bm k},ij}^\lambda f_{{\bm k},\lambda}^T \ .
\end{equation}
This decomposition makes sure that $\sigma_{\bm k}^{(1)}$, $\sigma_{{\bm k},i}^{(1)}$ and $\sigma^{(1)}_{{\bm k},\lambda}$ only depend on $f_{\bm k}^{S(1)}$, $f_{\bm k}^{V{(1)}}$ and $f_{{\bm k},\lambda}^{T{(1)}}$, respectively.  For example, 
\begin{equation}\label{eq:example}
    \sigma^{(1)}_{{\bm k},\lambda} \propto \epsilon_{{\bm k},ij, \lambda} \int {\rm d}\Omega_q \gamma^i \gamma^j f_{\bm k}^{(1)} = \epsilon_{{\bm k},ij}^\lambda \int {\rm d}\Omega_q \gamma^i \gamma^j \gamma^r \gamma^s \epsilon_{rs}^{\lambda'} f_{{\bm k},\lambda'}^{T(1)} \ ,
\end{equation}
which is consistent with the first order results shown in eq.~(\ref{eq:sigma1_t}). However, the decomposition could not be extended to the second order case. The reason that eq.~(\ref{eq:example}) holds is that in the $\eta'$-integral of $f_{\bm k}^{(1)}$, the index of $e$-exponential is $i\gamma_j k^j(\eta-\eta')$ and $k^i \epsilon_{{\bm k},ij,\lambda} = 0$. For the second order anisotropic stress, in the $\eta'$-integral the index of $e$-exponential is $i\gamma_j(k\eta-k_1\eta'-k_2\eta'')^j$ or $i\gamma_j(k\eta-k_2\eta'-k_1\eta'')^j$ and generally $k_1^i k_2^j \epsilon_{{\bm k},ij,\lambda}\neq 0$. Therefore, the terms contain only one or more $\gamma^i$ all contributed to the anisotropic stress. The terms, like $\gamma^{i_1}...\gamma^{i_n}\phi^{(1)}\psi^{(1)}, (n\neq 0, 2)$ are not $0$ after the angular integration and the effect of the operator $\epsilon_{ij}^\lambda$. They are omitted in previous work \cite{Mangilli:2008bw}. Here we have considered all of them. Refer to Sec.~\ref{sec:con} and \ref{sec:P_and_Q} for the detailed analyses.

\section{The transfer and kernel functions}\label{sec:trans_and_kernel}

\subsection{The transfer functions}
Expanding Einstein equation to the first order, we obtain the equations of the scalar and tensor perturbations,
\begin{align}
    -\phi_{\bm k}^{(1)} + \psi_{\bm k}^{(1)} &= \kappa a^2 \sigma_{\bm k}^{(1)} \ ,\label{eq:s_o1_1}\\ 
    2\mathcal{H}\left(3(c_s^2-\omega)\mathcal{H}\phi_{\bm k}^{(1)} + \phi^{{(1)}\prime}_{\bm k}+(2+3c_s^2)\psi_{\bm k}^{(1)\prime}\right)+2\psi_{\bm k}^{(1)\prime\prime} -k^2\phi_{\bm k}^{(1)} + k^2\psi_{\bm k}^{(1)} + 2c_s^2 k^2 \psi_{\bm k}^{(1)}&=\frac{\kappa a^2}{3}k^2 \sigma_{\bm k}^{(1)} \ ,\label{eq:s_o1_2}\\
    h_{\lambda, \bm k}^{(1)}{''} + 2\mathcal{H}h_{\lambda, \bm k}^{(1)}{'} + h_{\lambda, \bm k}^{(1)} &= 2\kappa a^2 \sigma_{\lambda,\bm k}^{(1)} \ ,\label{eq:t_o1_1}
\end{align}
where $\kappa \equiv 8\pi G$, $\omega$ is the pressure to energy-density ratio, and $c_s$ is the speed of sound.

Usually, we write the first order scalar and tensor perturbations with transfer functions,
\begin{equation}
    \psi_{\bm k}^{(1)}(\eta) \equiv \Phi_{\bm k} T_\psi (k\eta) \ , \quad  \phi_{\bm k}^{(1)}(\eta) \equiv \Phi_{\bm k} T_\phi (k\eta) \ , \quad h_{{\bm k},\lambda}^{(1)}(\eta) \equiv h^\mathrm{in}_{{\bm k},\lambda} \chi (k\eta) \ ,   
\end{equation}
where $\Phi_{\bm k}$ is the initial value originated from primordial curvature perturbation, $h^\mathrm{in}_{{\bm k},\lambda}$ is the initial value of the primordial gravitational wave mode, $T_\psi(k\eta)$, $T_\phi(k\eta)$ and $\chi(k\eta)$ are transfer functions of $\psi_{\bm k}$, $\phi_{\bm k}$ and $h_{{\bm k},\lambda}$, respectively.

\begin{figure}
    \centering
    \includegraphics[scale = 0.5]{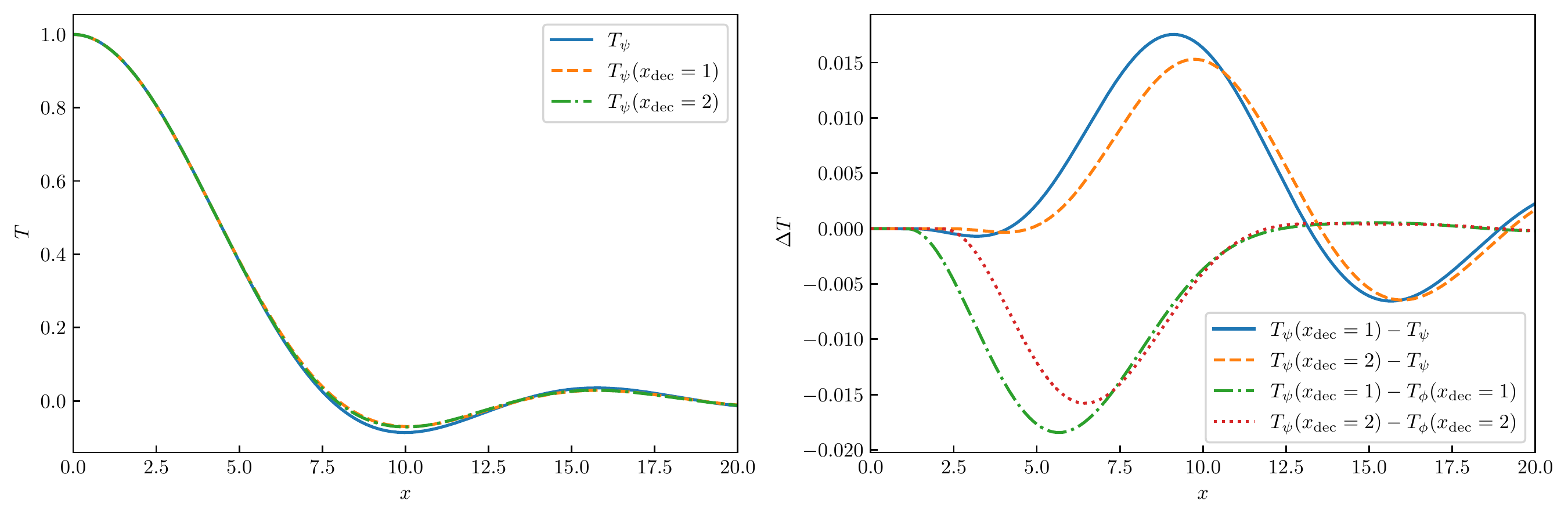}  
    \caption{The transfer functions of the first order scalar perturbations. Left panel shows the no damping $T_\psi$ (blue solid curve), damping $T_\psi$ for $x_\mathrm{dec}=1$ (orange dashed curve) and $x_\mathrm{dec}=2$ (green dot-dashed curve), where $x_\mathrm{dec} \equiv k \eta_\mathrm{dec}$. The horizontal axis represents $x\equiv k\eta$. Right panel gives the differences between the different transfer functions.}\label{fig:s_o1}
\end{figure} 

\begin{figure}
    \centering
    \includegraphics[scale = 0.5]{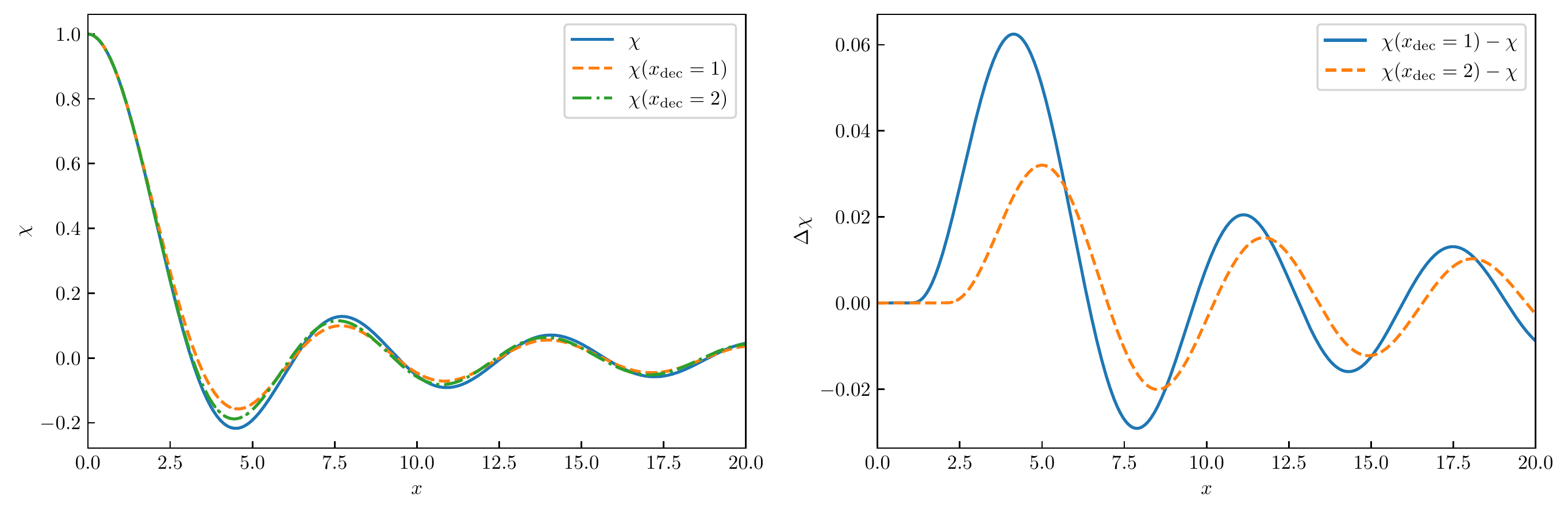}
    \caption{The transfer functions of the first order tensor perturbation. Left panel shows the no damping $\chi$ (blue solid curve), damping $\chi$ for $x_\mathrm{dec}=1$ (orange dashed curve) and $x_\mathrm{dec}=2$ (green dot-dashed curve). Right panel gives the differences between the damping and no damping transfer functions.}\label{fig:t_o1}
\end{figure}

The transfer functions of the first order scalar and tensor perturbations are shown in Figs.~\ref{fig:s_o1} and \ref{fig:t_o1}. The results of the first order tensor perturbations have been obtained by Weinberg earliest \cite{Weinberg:2003ur}. Later, Ref.~\cite{Stefanek:2012hj} gave the fitting formula of the transfer function $\chi(x)$. Here we have set the effective number of neutrinos $N_\mathrm{eff}=3.046$ and the gravitational waves evolve in radiation-dominated era. To study the neutrino effects to the perturbations of different frequencies, we give two damping transfer functions of $x_\mathrm{dec}=1,2$ in each figure, where $x_\mathrm{dec}\equiv k\eta_\mathrm{dec}$. The damping of the perturbations would only arise after the decoupling of the neutrinos. From the right panels of Figs.~\ref{fig:s_o1} and \ref{fig:t_o1} we obtain that the curves of $x_\mathrm{dec}=2$ are much closer to the no damping ones. It is consistent with eqs.~(\ref{eq:sigma1_s}) and (\ref{eq:sigma1_t}), in which the anisotropic stress induced by neutrinos are much smaller for high-frequency perturbations than low-frequency perturbations. On the other hand, Fig.~\ref{fig:s_o1} shows that for the one order scalar perturbations, neutrino makes $T_\psi$ and $T_\phi$ different with small $x$ (green dot-dashed and red dotted curves). When $x$ becomes larger, $T_\psi \rightarrow T_\phi$. This could be explained by eqs.~(\ref{eq:s_o1_1}). $\sigma_{\bm k}^{(1)}$ is higher order infinitesimal compared with $\psi$ and $\phi$ as $x \rightarrow \infty$. 

\subsection{The kernel functions}
Considering the anisotropic stress contributed by neutrinos, we present the equation of motion of the second order induced gravitational waves as 
\begin{equation}\label{eq:2o_h_eq0}
    h_{ij}^{(2)\prime\prime}+2\mathcal{H}h_{ij}^{(2)\prime}-\Delta h_{ij}^{(2)} = 4\Lambda_{ij}^{rs}(S_{1rs}^{(2)} + S_{2rs}^{(2)} + S_{3rs}^{(2)} - 2\kappa a^2\Pi_{rs}^{(1)}\psi^{(1)} + \frac{1}{2}\kappa a^2\Pi_{rs}^{(2)}) \ ,
\end{equation}
where $\Pi_{rs}^{(n)}$ is the $n$-order anisotropic stress induced by neutrinos. $S_{1rs}^{(2)}$, $S_{2rs}^{(2)}$, and $S_{3rs}^{(2)}$ denotes the sources contributed by $\psi^{(1)}$ ($\phi^{(1)}$), $\psi^{(1)}$ ($\phi^{(1)}$) and $h_{rs}^{(1)}$, and $h_{rs}^{(1)}$, respectively.  
$\Lambda_{ij}^{rs} \equiv \epsilon_{ij}^\lambda \epsilon^{rs}_\lambda$ denotes a transverse and traceless operator \cite{Chang:2020tji}. Here we have used \texttt{xpand} package \cite{Pitrou:2013hga} to expand the Einstein equation to the second order. The three sources are written as 

\begin{equation}
    \begin{split}
        S_{1rs}^{(2)} = &-\partial_r\phi^{(1)}\partial_s\phi^{(1)} + \partial_r\psi^{(1)}\partial_s\phi^{(1)} + \partial_r\phi^{(1)}\partial_s\psi^{(1)} - 3\partial_r\psi^{(1)}\partial_s\psi^{(1)} \\
        &+ \frac{4(\mathcal{H}\partial_r\phi^{(1)}+\partial_r\psi^{(1)\prime})(\mathcal{H}\partial_s\phi^{(1)}+\partial_s\psi^{(1)\prime})}{3(1+w)^2\mathcal{H}^2}+ \frac{4w(\mathcal{H}\partial_r\phi^{(1)} - \partial_r\psi^{(1)\prime})(\mathcal{H}\partial_s\phi^{(1)}+\partial_s\psi^{(1)\prime})}{3(1+w)^2\mathcal{H}^2}\\ 
        &- 2\phi^{(1)}\partial_r\partial_s\phi^{(1)} - 2\psi^{(1)}\partial_r\partial_s\psi^{(1)} \ ,
    \end{split}
\end{equation}

\begin{equation}
    \begin{split}
        S_{2rs}^{(2)} = &\frac{1}{2} \big( 2h_{rs}^{(1)}{''}\phi^{(1)} + 4\mathcal{H}h_{rs}^{(1)}{'}\phi^{(1)} - 4\mathcal{H}^2h_{rs}^{(1)}\phi^{(1)} + 4(1+3\omega)\mathcal{H}^2 h_{rs}^{(1)}\phi^{(1)} - 4\mathcal{H}h_{rs}^{(1)}\phi^{(1)}{'}\\
        &-12\mathcal{H}h_{rs}^{(1)}\psi^{(1)}{'} - 6h_{rs}^{(1)}\psi^{(1)}{''} + 2\psi^{(1)}\Delta h_{rs}^{(1)} - 2h_{rs}^{(1)}\Delta\phi^{(1)} - 4c_s^2 h_{rs}^{(1)}(3\mathcal{H}^2\phi^{(1)} + 3\mathcal{H}\psi^{(1)}{'} - \Delta\psi^{(1)})\\
        &+ 4h_{rs}^{(1)}\Delta\psi^{(1)} - 2h_r^{l(1)}\partial_l\partial_s\psi^{(1)} - 2h_s^{l(1)}\partial_l\partial_r\psi^{(1)} + \partial_l h_{rs}^{(1)}\partial^l \phi^{(1)} + 3\partial_l h_{rs}^{(1)}\partial^l\psi^{(1)} - \partial^l\phi^{(1)}\partial_r h_{sl}^{(1)}\\
        &- \partial^l\psi^{(1)}\partial_r h_{sl}^{(1)} - \partial^l\phi^{(1)}\partial_s h_{rl}^{(1)} - \partial^l\psi^{(1)}\partial_s h_{rl}^{(1)}) \big)\ ,
    \end{split}
\end{equation}

\begin{equation}
    \begin{split}
        S_{3rs}^{(2)} = &\frac{1}{4} \big( 2h_r^{l(1)\prime}h_{sl}^{(1)\prime} - 2h^{lm(1)}\partial_l\partial_m h_{rs}^{(1)} + 2h^{lm(1)}\partial_m\partial_r h_{sl}^{(1)}+2h^{lm(1)}\partial_m\partial_s h_{rl}^{(1)} \\
        &+ 2\partial_l h_{sm}^{(1)}\partial^m h_r^{l(1)} - 2\partial_l h_{sm}^{(1)}\partial^l h_r^{m(1)} - \partial_r h^{lm(1)}\partial_s h_{lm}^{(1)} - 2h^{lm(1)}\partial_r \partial_s h_{lm}^{(1)} \big) \ .
    \end{split}
\end{equation}

Define the transfer function $f_i\quad (i=1,2{\rm \uppercase\expandafter{\romannumeral1}},2{\rm \uppercase\expandafter{\romannumeral2}},3{\rm \uppercase\expandafter{\romannumeral1}}, 3{\rm \uppercase\expandafter{\romannumeral2}}, 3{\rm \uppercase\expandafter{\romannumeral3}}, 3{\rm \uppercase\expandafter{\romannumeral4}})$ of the three sources as
\begin{align}
    S_{1{\bm k},ij}^{(2)} = \int \frac{{\rm d}^3 k_1 {\rm d}^3 k_2}{(2\pi)^3} &\delta({\bm k}_1+{\bm k}_2-{\bm k})k_{1i}k_{2j}f_{1}\Phi_{{\bm k}_1}\Phi_{{\bm k}_2} \ ,\label{eq:S_1}\\
    S_{2{\bm k},ij}^{(2)} = \int \frac{{\rm d}^3 k_1 {\rm d}^3 k_2}{(2\pi)^3} &\delta({\bm k}_1+{\bm k}_2-{\bm k}) (\epsilon_{{\bm k}_1,ij}^{\lambda'}k^2 f_{2{\rm \uppercase\expandafter{\romannumeral1}}} + \epsilon_{{\bm k}_1,il}^{\lambda'}k_{1j}k_2^l f_{2{\rm \uppercase\expandafter{\romannumeral2}}})h_{{\bm k}_1, \lambda'}^\mathrm{in}\Phi_{{\bm k}_1} \ ,\label{eq:S_2}\\
    S_{3{\bm k},ij}^{(2)} = \int \frac{{\rm d}^3 k_1 {\rm d}^3 k_2}{(2\pi)^3} &\delta({\bm k}_1+{\bm k}_2-{\bm k}) \big(\delta^{rs}\epsilon_{{\bm k}_1,ir}^{\lambda'} \epsilon_{{\bm k}_2,js}^{\lambda''}k^2 f_{3{\rm \uppercase\expandafter{\romannumeral1}}} + \epsilon_{{\bm k}_2,ij}^{\lambda''} \epsilon_{{\bm k}_1}^{rs,\lambda'}k_r k_s f_{3{\rm \uppercase\expandafter{\romannumeral2}}} + k_{1i} \epsilon_{{\bm k}_2,jr}^{\lambda''} \epsilon_{{\bm k}_1}^{rs,\lambda'}k_{2s}f_{3{\rm \uppercase\expandafter{\romannumeral3}}} \notag\\
    &+ k_{1i}k_{2j}\epsilon_{{\bm k}_1,lm}^{\lambda'}\epsilon_{{\bm k}_2}^{lm,\lambda''}f_{3{\rm \uppercase\expandafter{\romannumeral4}}} + \epsilon_{{\bm k}_1,ir}^{\lambda'}\epsilon_{{\bm k}_2,js}^{\lambda''}k^r k^s f_{3{\rm \uppercase\expandafter{\romannumeral5}}} \big) h_{{\bm k}_1,\lambda'}^\mathrm{in} h_{{\bm k}_2,\lambda''}^\mathrm{in} \label{eq:S_3} \ .
\end{align}

Eq.~(\ref{eq:2o_sigma1}) can be rewritten as 
\begin{align}
    \kappa a^2 \sigma_{{\bm k},\lambda}^{(2)} = 2 \int\frac{{\rm d}^3 k_1 {\rm d}^3 k_2}{(2\pi)^3}& \delta({\bm k}_1+{\bm k}_2-{\bm k}) [\epsilon_{{\bm k},\lambda}^{ij}k_{1i}k_{2j}\mathcal{P}_1\Phi_{{\bm k}_1}\Phi_{{\bm k}_2} + (\epsilon_{{\bm k},\lambda}^{ij} \epsilon_{{\bm k}_1,ij}^{\lambda'} k^2 \mathcal{P}_{2{\rm \uppercase\expandafter{\romannumeral1}}} + \epsilon_{{\bm k},\lambda}^{ij} \epsilon_{{\bm k}_1,il}^{\lambda'}k_{1j}k_2^l \mathcal{P}_{2{\rm \uppercase\expandafter{\romannumeral2}}})\Phi_{{\bm k}_1}h_{{\bm k}_2,\lambda'}^\mathrm{in}\notag\\
    &+(\delta^{rs}\epsilon_{{\bm k},\lambda}^{ij}\epsilon_{{\bm k}_1,ir}^{\lambda'} \epsilon_{{\bm k}_2,js}^{\lambda''}k^2 \mathcal{P}_{3{\rm \uppercase\expandafter{\romannumeral1}}} + \epsilon_{{\bm k},\lambda}^{ij} \epsilon_{{\bm k}_2,ij}^{\lambda''} \epsilon_{{\bm k}_1}^{rs,\lambda'}k_r k_s \mathcal{P}_{3{\rm \uppercase\expandafter{\romannumeral2}}} + k_{1i}\epsilon_{{\bm k},\lambda}^{ij} \epsilon_{{\bm k}_2,jr}^{\lambda''} \epsilon_{{\bm k}_1}^{rs,\lambda'}k_{2s}\mathcal{P}_{3{\rm \uppercase\expandafter{\romannumeral3}}} \notag\\
    &+ \epsilon_{\bm{k},\lambda}^{ij} k_{1i}k_{2j}\epsilon_{{\bm k}_1,lm}^{\lambda'}\epsilon_{{\bm k}_2}^{lm,\lambda''}\mathcal{P}_{3{\rm \uppercase\expandafter{\romannumeral4}}} + \epsilon_{{\bm k},\lambda}^{ij}\epsilon_{{\bm k}_1,ir}^{\lambda'}\epsilon_{{\bm k}_2,js}^{\lambda''}k^r k^s \mathcal{P}_{3{\rm \uppercase\expandafter{\romannumeral5}}})h_{{\bm k}_1,\lambda'}^\mathrm{in} h_{{\bm k}_2,\lambda''}^\mathrm{in} ] \notag\\
    &- 2\kappa a^2 \rho_\nu^{(0)} \int_{x_\mathrm{dec}}^x {\rm d}x \left( \frac{j_2(x-x')}{(x-x')^2} h_{{\bm k},\lambda}^{(2)} \right)  \ .
\end{align}\label{eq:sigma_o2_decomposition}

Similarly, $h_{{\bm k},\lambda}^{(2)}$ can be written as 
\begin{equation}\label{eq:h_compose_of_K}
    \begin{split}
        h_{{\bm k},\lambda}^{(2)} = \int\frac{{\rm d}^3 k_1 {\rm d}^3 k_2}{(2\pi)^3}&\delta({\bm k}_1+{\bm k}_2-{\bm k})[\epsilon_{{\bm k},\lambda}^{ij}k_{1i}k_{2j}K_1\Phi_{{\bm k}_1}\Phi_{{\bm k}_2} + (\epsilon_{{\bm k},\lambda}^{ij} \epsilon_{{\bm k}_1,ij}^{\lambda'} k^2 K_{2{\rm \uppercase\expandafter{\romannumeral1}}} + \epsilon_{{\bm k},\lambda}^{ij} \epsilon_{{\bm k}_1,il}^{\lambda'}k_{1j}k_2^l K_{2{\rm \uppercase\expandafter{\romannumeral2}}})\Phi_{{\bm k}_1}h_{{\bm k}_2,\lambda'}^\mathrm{in}\\
        &+(\delta^{rs}\epsilon_{{\bm k},\lambda}^{ij}\epsilon_{{\bm k}_1,ir}^{\lambda'} \epsilon_{{\bm k}_2,js}^{\lambda''}k^2 K_{3{\rm \uppercase\expandafter{\romannumeral1}}} + \epsilon_{{\bm k},\lambda}^{ij} \epsilon_{{\bm k}_2,ij}^{\lambda''} \epsilon_{{\bm k}_1}^{rs,\lambda'}k_r k_s K_{3{\rm \uppercase\expandafter{\romannumeral2}}} + k_{1i}\epsilon_{{\bm k},\lambda}^{ij} \epsilon_{{\bm k}_2,jr}^{\lambda''} \epsilon_{{\bm k}_1}^{rs,\lambda'}k_{2s}K_{3{\rm \uppercase\expandafter{\romannumeral3}}}\\
        & + \epsilon_{\bm{k},\lambda}^{ij} k_{1i}k_{2j}\epsilon_{{\bm k}_1,lm}^{\lambda'}\epsilon_{{\bm k}_2}^{lm,\lambda''}K_{3{\rm \uppercase\expandafter{\romannumeral4}}} + \epsilon_{{\bm k},\lambda}^{ij}\epsilon_{{\bm k}_1,ir}^{\lambda'}\epsilon_{{\bm k}_2,js}^{\lambda''}k^r k^s K_{3{\rm \uppercase\expandafter{\romannumeral5}}})h_{{\bm k}_1,\lambda'}^\mathrm{in} h_{{\bm k}_2,\lambda''}^\mathrm{in}] \ .
    \end{split}
\end{equation}

In eqs.~(\ref{eq:S_1})-(\ref{eq:h_compose_of_K}), each second order quantity has been decomposed into three components, scalar-scalar, scalar-tensor and tensor-tensor coupling terms. Besides, according to the contraction form of the polarization tensor $\epsilon_{ij,\lambda}$ and the spatial momentum, each component is divided again, e.g., $K_2\rightarrow K_{2{\rm \uppercase\expandafter{\romannumeral1}}} + K_{2{\rm \uppercase\expandafter{\romannumeral2}}}$. 


Finally, we obtain the equation for the kernel function $K_i$ ($i=1,2{\rm \uppercase\expandafter{\romannumeral1}}, 2{\rm \uppercase\expandafter{\romannumeral2}}, 3{\rm \uppercase\expandafter{\romannumeral1}}, 3{\rm \uppercase\expandafter{\romannumeral2}}, 3{\rm \uppercase\expandafter{\romannumeral3}}, 3{\rm \uppercase\expandafter{\romannumeral4}}, 3{\rm \uppercase\expandafter{\romannumeral5}}$)
\begin{equation}\label{eq:2o_final}
    \begin{split}
        K_{i}''+2\mathcal{H}K_{i}'+k^2 K_{i} =& 4\bigg[f_{i} + \mathcal{Q}_i + \mathcal{P}_i - 2\kappa a^2\rho_\nu^{(0)}\int_{x_\mathrm{dec}}^x{\rm d}x'\left(\frac{j_2(x-x')}{(x-x')^2}kK_i'\right)\bigg] \ ,\\
    \end{split}
\end{equation}
where $\mathcal{Q}_i$ comes from the terms composed of $\Pi_{ij}^{(1)}$ and $\psi^{(1)}$. $\mathcal{P}_i$ and the following integral term come from $\Pi_{ij}^{(2)}$. The explicit expressions of $\mathcal{P}_i$ and $\mathcal{Q}_i$ are shown in \ref{sec:P_and_Q}.

\begin{figure}
    \centering
    \includegraphics[scale = 0.5]{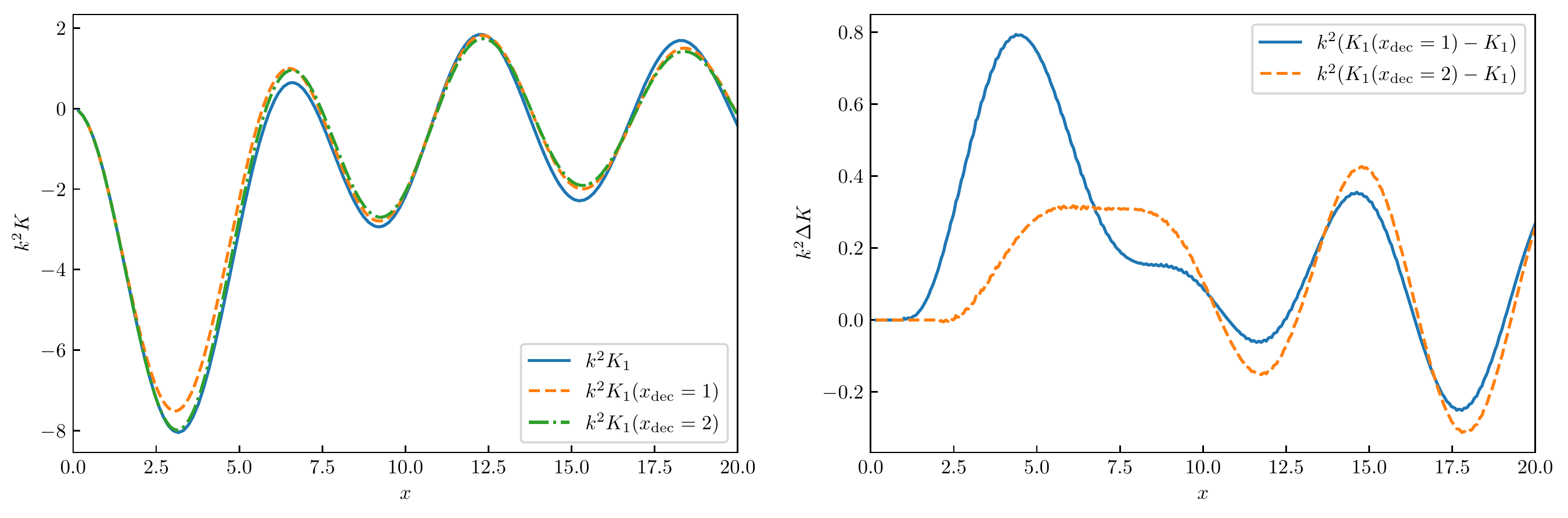}
    \caption{The kernel functions $K_1(u,v,x)$ of the second order tensor perturbation. We have set $u=v=1$ in this figure. Left panel shows the no damping $K_1$, damping $K_1$ (blue solid curve) for $x_\mathrm{dec}=1$ (orange dashed curve) and $x_\mathrm{dec}=2$ (green dot-dashed curve). Right panel gives the differences between the damping and no damping kernel functions.}\label{fig:ss_o2}
\end{figure}

\begin{figure}
    \centering
    \includegraphics[scale = 0.5]{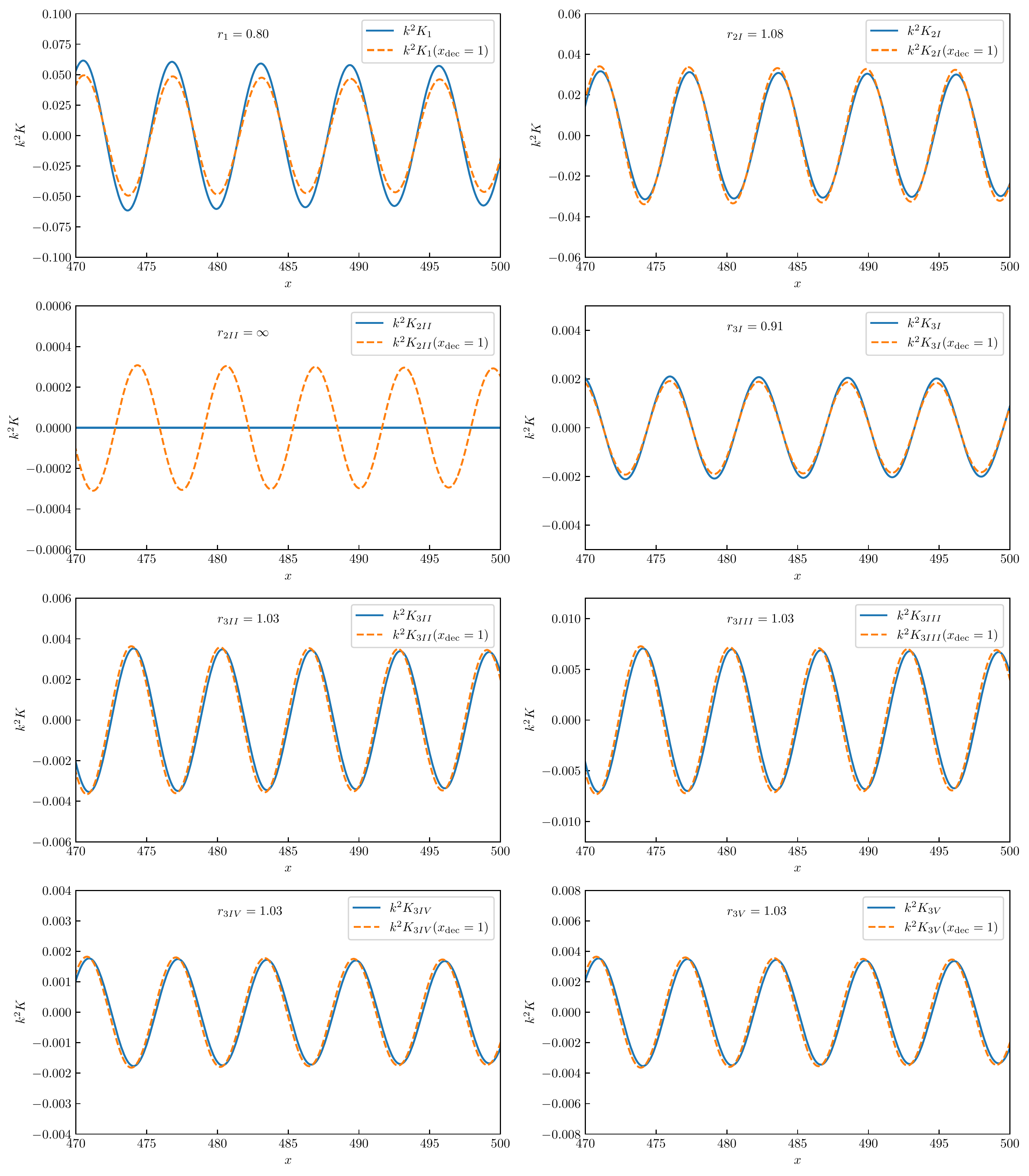}
    \caption{The kernel functions $K_i(u,v,x)\ ($i=1,2{\rm \uppercase\expandafter{\romannumeral1}}, 2{\rm \uppercase\expandafter{\romannumeral2}}, 3{\rm \uppercase\expandafter{\romannumeral1}}, 3{\rm \uppercase\expandafter{\romannumeral2}}, 3{\rm \uppercase\expandafter{\romannumeral3}}, 3{\rm \uppercase\expandafter{\romannumeral4}}, 3{\rm \uppercase\expandafter{\romannumeral5}}$)$ of the second order tensor perturbation. We have set $u=v=1$. For each mode the no damping (blue solid curve) and damping ($x_\mathrm{dec}=1$, orange dashed curve) kernel functions are given. The ratios of the amplitudes of two kernel functions $r$ are shown in each panel.}\label{fig:kernel_function_far}
\end{figure}

\begin{figure}
    \centering
    \includegraphics[scale = 0.6]{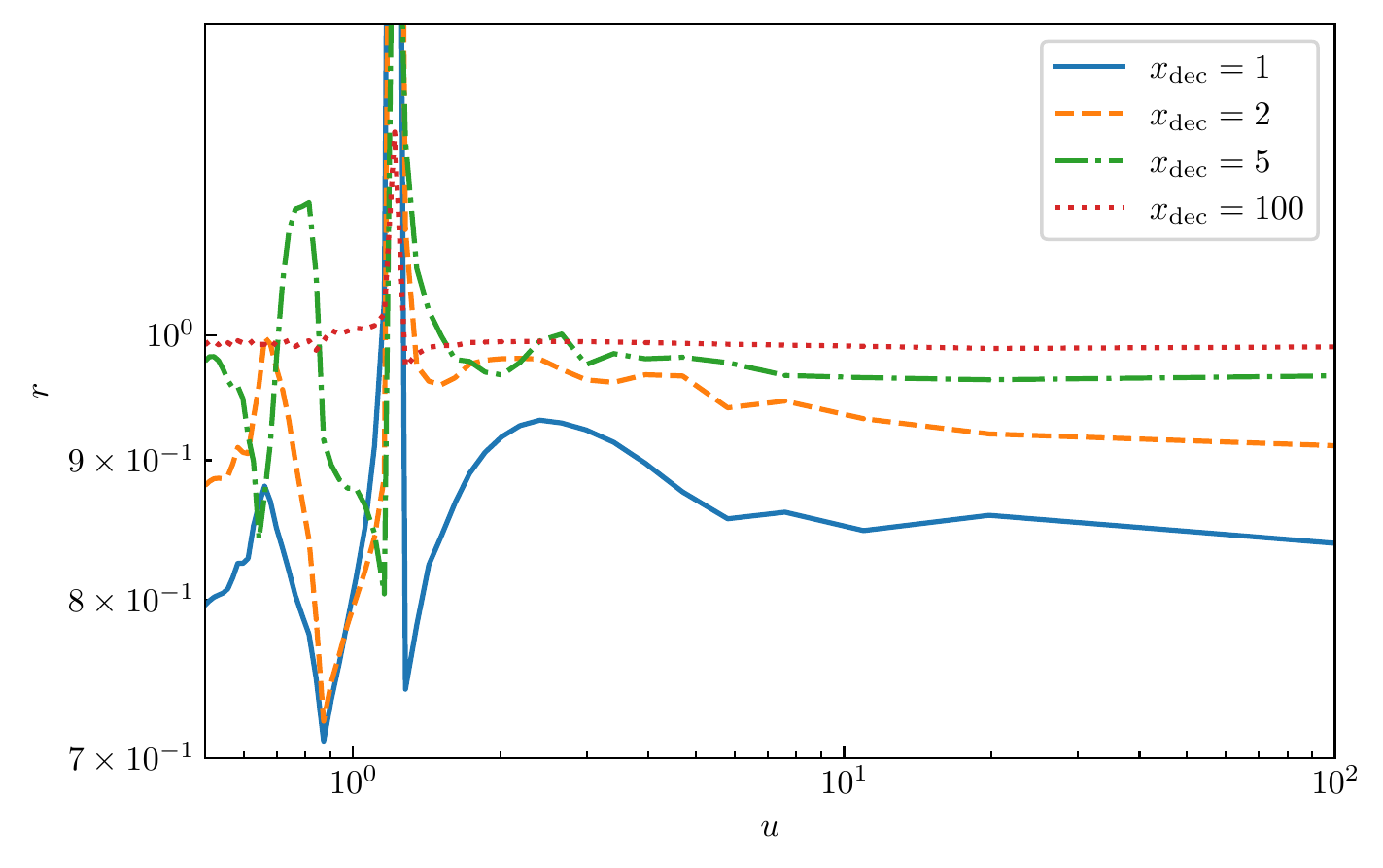}
    \caption{The ratio $r_1$ of the damping to the no damping second order kernel functions $K_1$ near $x\sim 500$. Here we have set $u=v$. Four kernel functions of different frequencies are shown.}\label{fig:ratio}
\end{figure}

It should be clarified that normally $K_i$ are the functions of $|{\bm k}_1|$, $|{\bm k}_2|$ and $x$. Define $u\equiv |{\bm k}_1|/|{\bm k}|$ and $v\equiv |{\bm k}_2|/|{\bm k}|$. Then $K_i$ could be written as $K_i(u,v,x)$. Eq.~(\ref{eq:2o_final}) shows that $K_i$ also depend on $x_\mathrm{dec}$ considering the neutrinos. When $u$, $v$ have been fixed, we denote the function $K_i(x)$ as $K_i(x_\mathrm{dec})(x)$ with no ambiguity. For example, in Fig.~\ref{fig:ss_o2} we give the kernel function $K_1$ in radiation-dominated era for small $x$. The damping gravitational waves of $x_\mathrm{dec}=1$ and $x_\mathrm{dec}=2$ are shown. For the mode of $x_\mathrm{dec}=2$, as it enter the horizon, neutrinos have not decoupled. $\mathcal{Q}_i$, $\mathcal{P}_i$ and the integral term in the right hand of eq.~(\ref{eq:2o_final}) keep $0$ until $x=x_\mathrm{dec}=2$. After this moment the neutrinos decouple and affect the propagation of the gravitational wave. 

In this paper we study the energy density of the gravitational waves from a monochromatic curvature perturbation. In this case only the behaviors of the kernel functions well inside the horizon ($x\rightarrow \infty$) make sense. In Fig.~\ref{fig:kernel_function_far}, we give the kernel functions $K_i$ ($i=1,2{\rm \uppercase\expandafter{\romannumeral1}}, 2{\rm \uppercase\expandafter{\romannumeral2}}, 3{\rm \uppercase\expandafter{\romannumeral1}}, 3{\rm \uppercase\expandafter{\romannumeral2}}, 3{\rm \uppercase\expandafter{\romannumeral3}}, 3{\rm \uppercase\expandafter{\romannumeral4}}, 3{\rm \uppercase\expandafter{\romannumeral5}}$) between $x=470\sim500$ with $u=v=1$. In each panel of Fig.~\ref{fig:kernel_function_far}, the no damping ($K_i$) and damping ($K_i(x_\mathrm{dec}=1)$) kernel functions are given. Notice that  $K_i = A\frac{\sin(x+\delta)}{x}$ as $x\rightarrow \infty$, which is consistent with the analytic formulas given by \cite{Kohri:2018awv}. The free-streaming neutrinos change the amplitudes and phases of $K_i$. As a result, we could define $\lim_{x\rightarrow \infty}K_i(x_\mathrm{dec}=1) \equiv r_i A\frac{\sin(x+\delta+\delta_i)}{x}$, where the ratios $r_i$ have shown in Fig.~\ref{fig:kernel_function_far}. We find that the kernel function $K_1$ has been damped to $80\%$ well inside the horizon for $u=v=1$, which is most affected. Moreover, eqs.~(\ref{eq:s_o1_1}) and (\ref{eq:f22}) imply that neutrinos make a distinction between two scalar perturbations, and thus $f_{2{\rm \uppercase\expandafter{\romannumeral2}}}$, $K_{2{\rm \uppercase\expandafter{\romannumeral2}}}$ become nonzero considering neutrinos. The kernel function $K_{3{\rm \uppercase\expandafter{\romannumeral1}}}$ is damped to $90\%$. For the rest kernel functions, they are enhanced slightly. 
Shown in eqs.~(\ref{eq:f32}-\ref{eq:f35}), $f_{3{\rm \uppercase\expandafter{\romannumeral2}}} \propto f_{3{\rm \uppercase\expandafter{\romannumeral3}}} \propto f_{3{\rm \uppercase\expandafter{\romannumeral4}}} \propto f_{3{\rm \uppercase\expandafter{\romannumeral5}}}$. In addition, as $x\rightarrow\infty$, $\mathcal{P}_i + \mathcal{Q}_i \ll$ "the integral term" in the right hand of eq.~(\ref{eq:2o_final}). Therefore, as the solutions of the eq.~(\ref{eq:2o_final}), $K_{3{\rm \uppercase\expandafter{\romannumeral2}}} \propto K_{3{\rm \uppercase\expandafter{\romannumeral3}}} \propto K_{3{\rm \uppercase\expandafter{\romannumeral4}}} \propto K_{3{\rm \uppercase\expandafter{\romannumeral5}}}$ when $x\rightarrow\infty$. As a result, we obtain $r_{3{\rm \uppercase\expandafter{\romannumeral2}}} = r_{3{\rm \uppercase\expandafter{\romannumeral3}}} = r_{3{\rm \uppercase\expandafter{\romannumeral4}}} = r_{3{\rm \uppercase\expandafter{\romannumeral5}}}$, which is shown in Fig.~\ref{fig:kernel_function_far}.

The theory of the standard slow-roll inflation \cite{book:15453} and recent observations \cite{Planck:2018vyg} indicate a tiny tensor-to-scalar ratio. Since the kernel functions $|K_{2i}|$, $|K_{3j}| \cancel{\gg} |K_1|$ ($i={\rm \uppercase\expandafter{\romannumeral1}}, {\rm \uppercase\expandafter{\romannumeral2}}$, $j={\rm \uppercase\expandafter{\romannumeral1}}\sim{\rm \uppercase\expandafter{\romannumeral5}}$), in this paper we could neglect the primordial tensor power spectrum safely. We only give eq.~(\ref{eq:2o_final}) and examples in Fig.~\ref{fig:kernel_function_far} of the kernel functions with the contributions of the first order tensor perturbations, which may be used for future study.

The intensity of the damping effect is dependent on the frequency $k$ of the gravitational wave and the parameters $u$ and $v$. We give illustrations in Fig.~\ref{fig:ratio}. It is shown that the free-streaming neutrinos damped the gravitational waves with lower frequencies more significantly. It is because that compared with higher frequency mode, lower frequency mode influenced by the neutrinos earlier. $K_1(x_\mathrm{dec}=1)$ has been suppressed to $71\%$ mostly at $u=v\simeq 0.87$. On the contrary, $K_1(x_\mathrm{dec}=100)$ influenced by neutrinos slightly. In addition, we have $K_1=0$ when $u=v=\sqrt{6}/2$ for the no damping kernel function \cite{Kohri:2018awv}. The effect of neutrinos changes this zero point and make the ratio here become infinite.

\begin{figure}
    \centering
    \includegraphics[scale = 0.6]{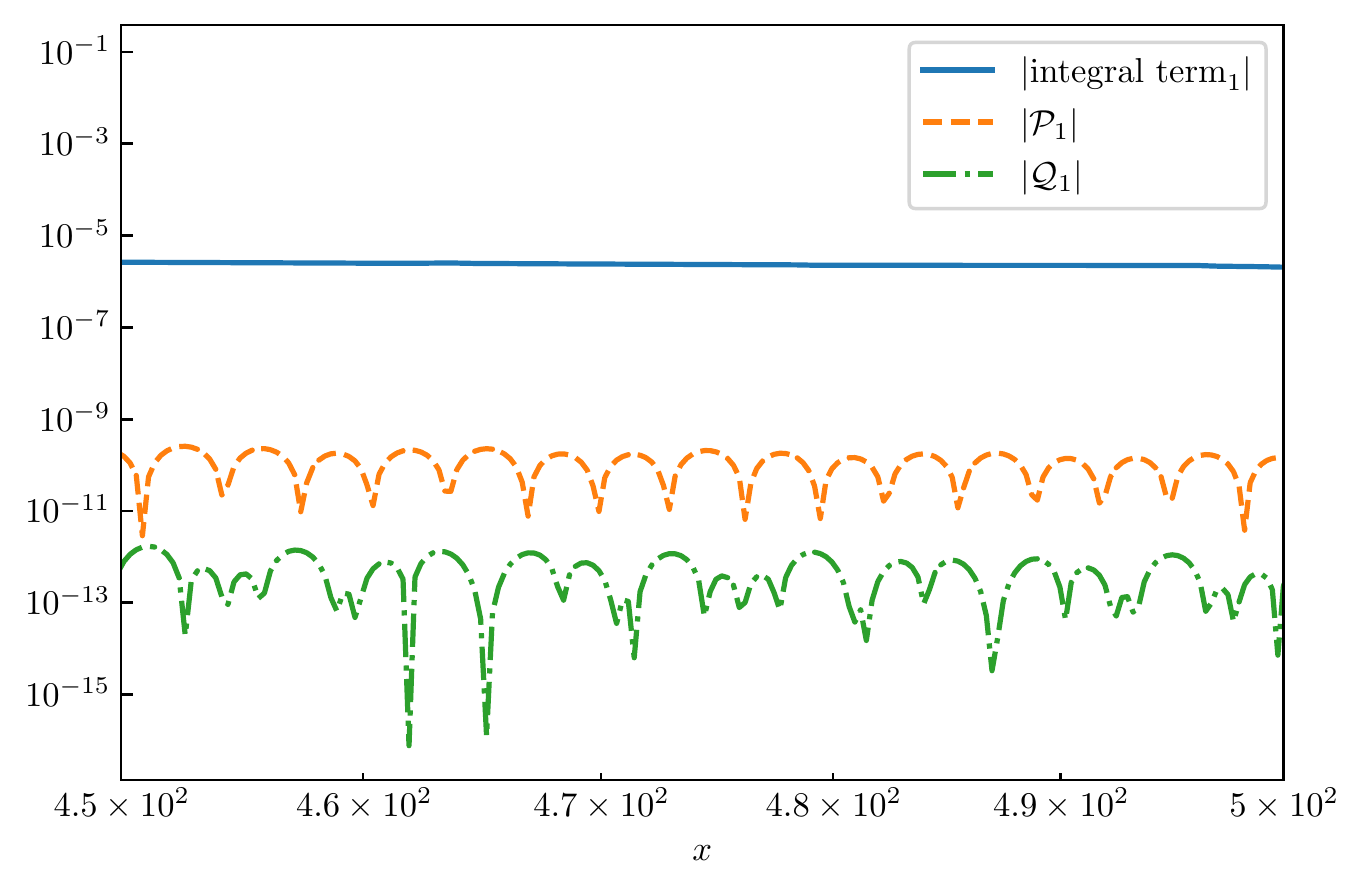}
    \caption{The comparison of the integral term (blue solid curve), $\mathcal{P}_1$ (orange dashed curve) and $\mathcal{Q}_1$ (green dot-dashed curve) in the right hand of eq.~(\ref{eq:2o_final}). Here we have set $u=v=1$.}\label{fig:compare_Q_P}
\end{figure}

In the last of this section we show the comparison of the integral term, $\mathcal{P}_1$ and $\mathcal{Q}_1$ in the right hand of eq.~(\ref{eq:2o_final}) well inside the horizon in Fig.~\ref{fig:compare_Q_P}. We find that for large $x$, $|\mathrm{the\ integral\ term}| \gg \mathcal{P}_1 + \mathcal{Q}_1$. Thus, we obtain that the contributions of neutrinos are mainly from the second order anisotropic stress $\Pi^{(2)}_{ij}$, rather than the couple of the first order scalar perturbation and anisotropic stress $\psi^{(1)}\Pi^{(1)}_{ij}$ (See eq.~(\ref{eq:2o_h_eq0})).

\section{The energy density spectrum}\label{sec:energy_spectrum}

As we mentioned before, the primordial tensor power spectrum $\mathcal{P}_h$ is neglected. In this section, we focus on the neutrino effect on the scalar induced gravitational waves, and give the energy density spectrum. The power spectrum $\mathcal{P}_h^{(2)}$ of the second order induced gravitational wave is defined as
\begin{equation}
    \left\langle h_{\bm k}^{\lambda(2)}(\eta) h_{{\bm k}'}^{\lambda{'}(2)}(\eta) \right\rangle = \delta^{\lambda\lambda'}\delta(k+k')\frac{2\pi^2}{k^3}\mathcal{P}_h^{(2)} (\eta,k) \ ,
\end{equation}
where $\left\langle h_{\bm k}^{\lambda(2)}(\eta) h_{{\bm k}'}^{\lambda{'}(2)}(\eta) \right\rangle $ is the two-point correlation function of $h_{\bm k}^{\lambda(2)}(\eta)$. Then we obtain \cite{Kohri:2018awv}
\begin{equation}
    \mathcal{P}_h^{(2)} = \frac{1}{4}\int_0^\infty {\rm d} v \int_{|1-v|}^{1+v} {\rm d}u \left\{k^4 \left(\frac{4v^2-(1+v^2-u^2)^2}{4uv}\right)^2 K_1^2(u,v,x)\mathcal{P}_\Phi(ku)\mathcal{P}_\Phi(kv)\right\} \ .
\end{equation}
Here, only the scalar induced part, i.e., $K_1$ in eq.~(\ref{eq:h_compose_of_K}) is considered. $\mathcal{P}_\Phi$ is the power spectrum of the first order scalar perturbation 
\begin{equation}
    \left\langle \Phi_{{\bm k}}\Phi_{{\bm k}'}  \right\rangle = \delta(k+k')\frac{2\pi^2}{k^3}\mathcal{P}_\Phi(k) \ .
\end{equation}
The fraction of the gravitational wave energy density per logarithmic wavelength is given by
\begin{equation}
    \Omega(\eta, k)= \frac{1}{24}\left(\frac{k}{a(\eta)H(\eta)}\right)^2\mathcal{P}_h^{(2)}(\eta,k) \ .
\end{equation}
Here, we consider the monochromatic curvature perturbations,
\begin{equation}
    \mathcal{P}_\Phi (k) = A k_* \delta(k-k_*) \ ,
\end{equation}
where $A$ denotes the overall normalization and $k_*$ is the location where the spectrum has a delta peak.

Finally, we obtain
\begin{equation}\label{eq:Omega}
    \Omega = \lim_{x\rightarrow\infty} \frac{A^2}{96}\frac{x^2}{\tilde{k}^2}\left(1-\frac{\tilde{k}^2}{4}\right)^2 k^4 K_1^2(\frac{1}{\tilde{k}}, \frac{1}{\tilde{k}}, x) \theta(2-\tilde{k})\theta(\tilde{k}) \ ,
\end{equation}
where $\tilde{k}=k/k_*$, and $\theta$ denotes the Heaviside step function.

\begin{figure}
    \centering
    \includegraphics[scale = 0.5]{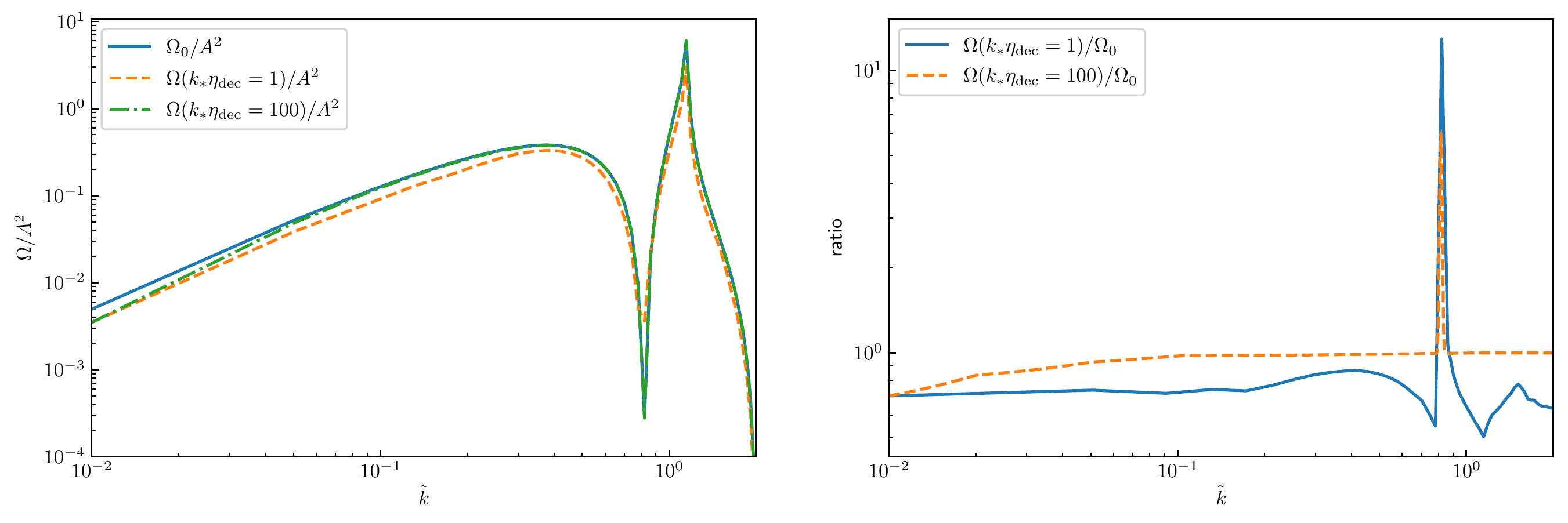}
    \caption{The energy density of the second order induced gravitational waves. Left panel shows the no damping energy density $\Omega_0$ (blue solid curve), damping energy density $\Omega(k_\ast \eta_\mathrm{dec}=1)$ and $\Omega(k_\ast \eta_\mathrm{dec}=100)$. Right panel gives the ratio $\Omega(k_\ast \eta_\mathrm{dec}=1)/\Omega_0$ and $\Omega(k_\ast \eta_\mathrm{dec}=100)/\Omega_0$.
    }\label{fig:ps}
\end{figure}

Considering the damping effect of the free-streaming neutrinos, we show the modified energy density spectra of the second order gravitational waves in Fig.~\ref{fig:ps}. A significant difference between the modified spectra and the no damping spectra is that the shapes of the former depend on the choices of the frequency $k_*$. Shown in Fig.~\ref{fig:ps}, the two spectra of $k_\ast \eta_\mathrm{dec}=1$ and $k_\ast \eta_\mathrm{dec}=100$ have an obvious difference. The right panel of Fig.~\ref{fig:ps} gives the ratio between the damping spectra $\Omega(x_\mathrm{dec})$ to the no damping spectrum $\Omega_0$. The curve of $\Omega(k_\ast \eta_\mathrm{dec}=1)/\Omega_0$ is below the curve of $\Omega(k_\ast \eta_\mathrm{dec}=100)/\Omega_0$, which means that the damping effect is more significant for former. There are peaks at $\tilde{k} = 2/\sqrt{6}$, which corresponds with the infinites at $u=\sqrt{6}/2$ in Fig.~\ref{fig:ratio}. We find that for the density spectrum of $k_\ast \eta = 100$, the neutrino could suppress it to $70\%$ at $\tilde{k}=0.01$. The damping effect gradually weakens with the increase of $\tilde{k}$ and nearly vanishes at $\tilde{k}=0.1$ and above. As a result, the logarithmic slope $n$ in the infrared region, which is defined as $\Omega \propto \tilde{k}^n$ at small $\tilde{k}$ \cite{Yuan:2019wwo}, jumps from $1.54$ to $1.63$ at $\tilde{k} = 0.01$.


\section{Conclusion and Discussion}\label{sec:con}

\begin{figure}
    \centering
    \includegraphics[scale = 0.5]{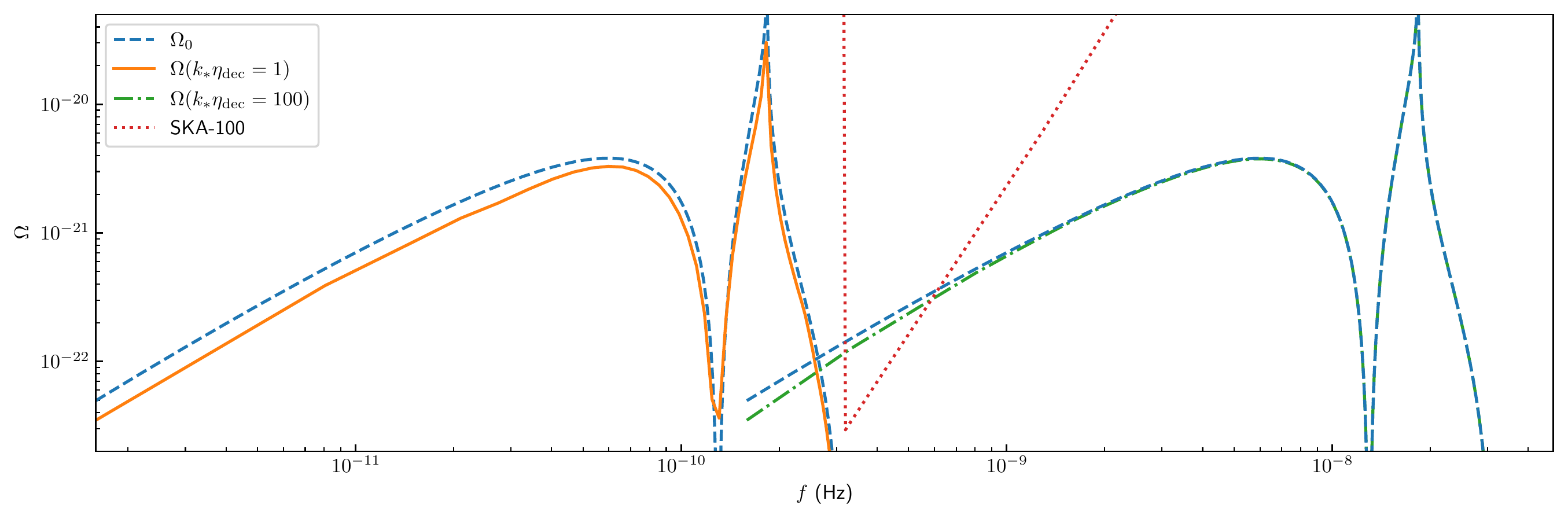}
    \caption{The energy density curves with taking into account of the effect of neutrinos and the sensitivity curve of SKA-100 \cite{Kuroda:2015owv}. Here, SKA-100 denotes the observations of $100$ pulsars with $20$ns timing accuracy for a $100$ year time span by \ac{SKA}. We have set $A=10^{-8}$.}\label{fig:ps2}
\end{figure}

Notice for the density spectrum of $k_\ast \eta_\mathrm{dec} = 100$, i.e., $k_\ast \sim 100 $nHz, the modified energy spectrum have the possibility of being probed by the \ac{PTA} in the future. In Fig.~\ref{fig:ps2}, we show the current density spectra $\Omega_\mathrm{current} = \Omega_R \Omega$ \cite{Espinosa:2017sgp, Yuan:2019udt} taking into account of the effect of neutrinos and the sensitivity curve of \ac{SKA} \cite{Kuroda:2015owv}. Here we have set $A=10^{-8}$. $\Omega_R$ is the density parameter of radiation at present. $f=k/2\pi$ is the frequency of the gravitational waves. Although the free-streaming neutrinos suppress $\Omega(k_\ast\eta_\mathrm{dec}=1)$ more significantly, unfortunately, it locates far below the frequency of nHz and cannot be detected by \ac{SKA}. For $\Omega(k_\ast\eta_\mathrm{dec}=100)$, \ac{SKA} may have a precise detection and figure out the neutrino effect on the gravitational waves. 

\begin{figure}
    \centering
    \includegraphics[scale = 0.6]{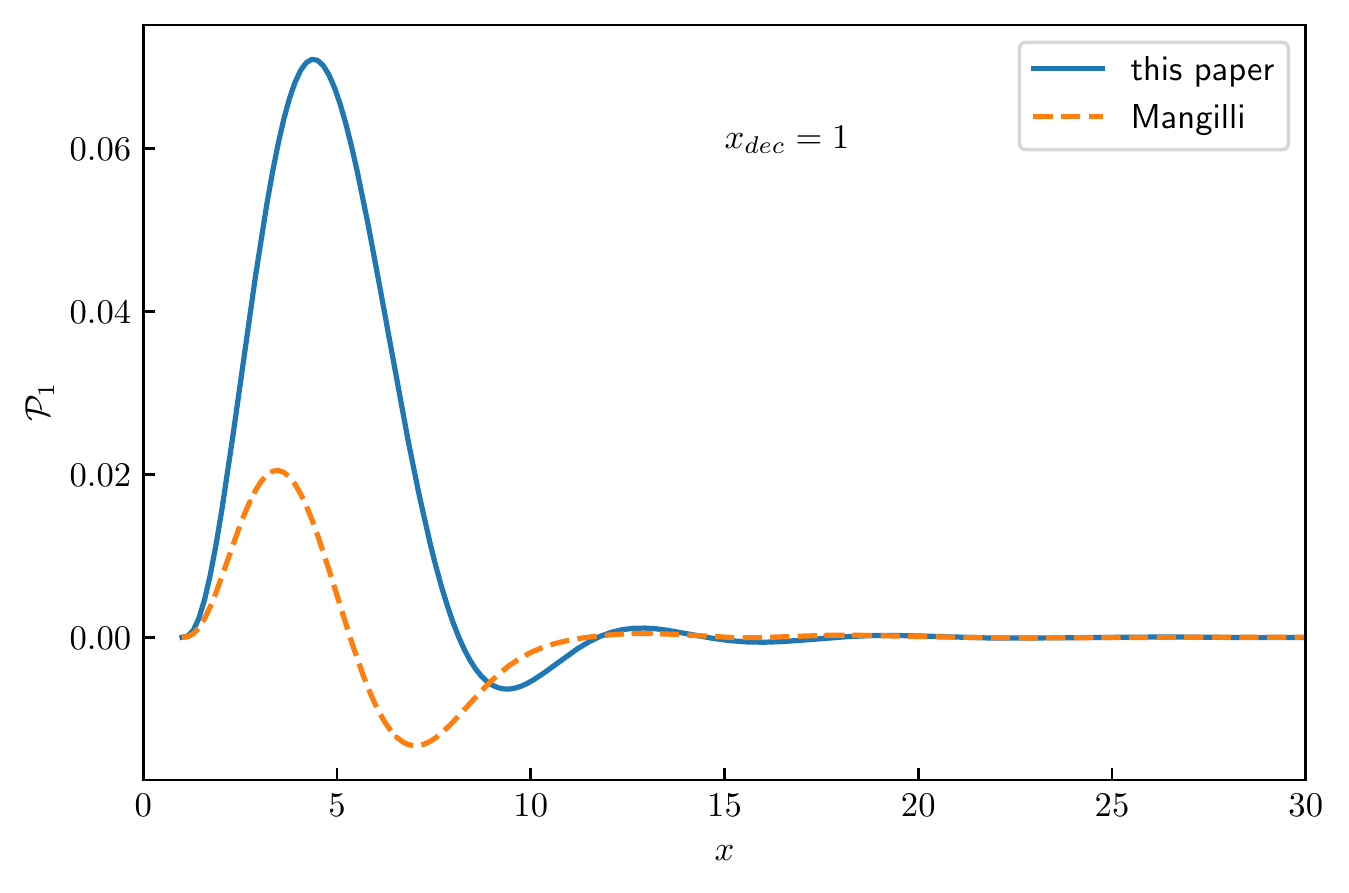}
    \caption{The comparison of $\mathcal{P}_1$ between our result and Mangilli's. Here we have set $x_\mathrm{dec}=1$ and $u=v=1$.}\label{fig:P1_comparison}
\end{figure}

In Sec.~\ref{cal:2nd} we have mentioned that we considered the integrations of the terms like $\gamma^{i_1}...\gamma^{i_n}\phi^{(1)}\psi^{(1)}$, $(n\neq 0, 2)$, which make a difference with Ref.~\cite{Mangilli:2008bw} in the second order anisotropic stress $\Pi_{ij}^{(2)}$. More precisely, this difference finally appears on $\mathcal{P}_1$ in eq.~(\ref{eq:2o_final}). Fig.~\ref{fig:P1_comparison} gives us the comparison between our $\mathcal{P}_1$ and Ref.~\cite{Mangilli:2008bw}. It shows that the two results have obvious difference at small $x$, which could affect $K_1$ at small $x$. As $x\rightarrow \infty$, since the two $\mathcal{P}_1$'s are of the same order and $|\mathrm{the\ integral\ term}| \gg \mathcal{P}_1 + \mathcal{Q}_1$, the kernel function $K_1$ well inside the horizon are the same.

In this paper, we analyzed the effect of the free-streaming neutrinos to the second order induced gravitational waves. We give the first order transfer function and the second order kernel functions considering the neutrinos. For the reason that the gravitational waves with lower frequencies influenced by the neutrinos for much longer times, they are damped more significantly than the higher frequency modes. Finally, we give the energy density spectrum of $k_\ast \sim 100$nHz. It has been damped to $70\%$ at $\tilde{k}=0.01$. As $\tilde{k}$ increases, the damping effect gradually vanishes, which enlarges the logarithmic slope $n$ in the infrared region. These effects may be examined by \ac{PTA} in the future.

\begin{acknowledgements}
We thank Prof. S. Wang and Dr. Q.H. Zhu for useful discussions. This work has been funded by the National Nature Science Foundation of China under grant No. 12075249 and 11690022, and the Key Research Program of the Chinese Academy of Sciences under Grant No. XDPB15.
\end{acknowledgements}

\appendix
\section{The Second Order Quantities}\label{sec:geodesic}
The temporal component of the geodesic equation is written as
\begin{equation}\label{eq:geo_0}
    \frac{{\rm d}P^0}{{\rm d}\lambda} = P^0\frac{{\rm d}P^0}{{\rm d}t} = -\Gamma^0_{\alpha\beta}P^\alpha P^\beta \ .
\end{equation}
Using eqs.~(\ref{eq:metric_t}) and (\ref{eq:geo_0}), we obtain
\begin{equation}\label{eq:dP^0/dt}
    \begin{split}
        \frac{{\rm d}P^0}{{\rm d}t} =& P^0\bigg( -\frac{\dot{a}}{a} - \partial_t \phi^{(1)} - \frac{2}{a}\gamma^i\partial_i\psi^{(1)} + \partial_t\psi^{(1)} - \frac{1}{2}\gamma^r \gamma^s \partial_t h_{rs}^{(1)} + \frac{1}{2}\bigg( 4\phi^{(1)}\partial_t\phi^{(1)} + \frac{4}{a}\phi^{(1)}\gamma^i\partial_i\phi^{(1)}\\
        &-\frac{4}{a}\psi^{(1)}\gamma^i\partial_i\phi^{(1)} + \frac{2}{a}\gamma^r\gamma^s h^{(1)}_{rs}\gamma^i\partial_i\phi^{(1)} + 4\psi^{(1)}\partial_t\psi^{(1)} - 2\psi^{(1)}\gamma^i\gamma^j\partial_t h_{ij}^{(1)} - 2\gamma^i\gamma^j h_{ij}^{(1)}\partial_t\psi^{(1)}\\
        &+ \gamma^i\gamma^j\gamma^r\gamma^s h_{ij}^{(1)} \partial_t h_{rs}^{(1)} - \partial_t\phi^{(2)} + \partial_t\psi^{(2)} - \frac{1}{2}\gamma^i\gamma^j\partial_t h_{ij}^{(2)} - \frac{2}{a}\gamma^i\partial_i\phi^{(2)} + \frac{1}{a}\gamma^i\gamma^j e_i\partial_j V^{(2)}\\
        &- \frac{2\dot{a}}{a}\gamma^i e_i V^{(2)} \bigg) \bigg) \ .
    \end{split}
\end{equation}

Considering that in the Boltzmann equation eq.(\ref{eq:Boltzmann_equation}), the comoving three momentum and comoving time $\eta$ are used. By using eqs.~(\ref{eq:p^2}) and (\ref{eq:p^i}) we can rewrite eq.~(\ref{eq:dP^0/dt}) as
\begin{equation}\label{eq:dq/deta}
    \begin{split}
        \frac{{\rm d}q}{{\rm d}\eta} =& q\bigg(-\gamma^i\partial_i\phi^{(1)} + \partial_\eta\psi^{(1)} - \frac{1}{2}\gamma^i\gamma^j\partial_\eta h_{ij}^{(1)} + \bigg( -\psi^{(1)}\gamma^i\partial_i\phi^{(1)} + \phi^{(1)}\partial_\eta\phi^{(1)} + 4\phi^{(1)}\gamma^i\partial_i\phi^{(1)}\\
        &+ \frac{1}{2}\gamma^r\gamma^s h_{rs}^{(1)}\gamma^i\partial_i\phi^{(1)} + 2\psi^{(1)}\partial_\eta\psi^{(1)} - \psi^{(1)}\gamma^i\gamma^j\partial_\eta h^{(1)}_{ij} - \gamma^i\gamma^j h^{(1)}_{ij}\partial_\eta \psi^{(1)} + \frac{1}{2}\gamma^i\gamma^j\gamma^r\gamma^s h_{ij}^{(1)} \partial_\eta h_{rs}^{(1)}\\ 
        &+ \frac{1}{2}\partial_\eta \psi^{(2)} - \frac{1}{4}\gamma^i \gamma^j \partial_\eta h^{(2)}_{ij} - \frac{1}{2}\gamma^i\partial_i\phi^{(2)} + \frac{1}{2}\gamma^i\gamma^j e_i \partial_j V^{(2)} - \mathcal{H}V^{(2)}e_i\gamma^i \bigg) \bigg) \ .
    \end{split}
\end{equation}
Similarly, using the spatial component of the geodesic equation, we could obtain $\frac{{\rm d}\gamma^i}{{\rm d}\eta}$ to the first order
\begin{equation}\label{eq:dgamma^i/deta}
    \begin{split}
        \frac{{\rm d}\gamma^i}{{\rm d}\eta} =& -\partial^i\phi^{(1)} - \partial^i\psi^{(1)} + \gamma^i\gamma^j\partial_j\phi^{(1)} + \gamma^i\gamma^j\partial_j\psi^{(1)} - 2\mathcal{H}\delta^{ij}\gamma^l h_{lj}^{(1)} - \delta^{ij}\gamma^l\partial_\eta h_{lj}^{(1)} - \delta^{ij}\gamma^k\gamma^l\partial_k h_{lj}^{(1)}\\
        &+ \frac{1}{2}\gamma^r\gamma^s\partial^i h^{(1)}_{rs} + 2\mathcal{H}\gamma_j h^{(1)ij} + \gamma^i\gamma^r\gamma^s\partial_\eta h^{(1)}_{rs} + \frac{1}{2}\gamma^i\gamma^j\gamma^r\gamma^s\partial_j h_{rs}^{(1)} \ .
    \end{split} 
\end{equation}
Following eqs.~(\ref{eq:metric_t}) and the constraint condition $g_{\mu\nu}P^\mu P^\nu=0$, we obtain
\begin{equation}\label{eq:dx^i/deta}
    \frac{{\rm d}x^i}{{\rm d}\eta} = a\frac{{\rm d}x^i}{{\rm d}t} = a\frac{P^i}{P^0} = \left(1+\phi^{(1)}+\psi^{(1)}-\frac{1}{2}\gamma^r\gamma^s h_{rs}^{(1)}\right)\gamma^i \ .
\end{equation}
Substituting eqs.~(\ref{eq:dq/deta})-(\ref{eq:dx^i/deta}) into eq.~(\ref{eq:Boltzmann_equation}), the explicit form of the Boltzmann equation to the second order is obtained.

The components of the right hand of eq.~(\ref{eq:definition_of_A}) can be written as
\begin{align}
    \left(\frac{{\rm d}x^i}{{\rm d}\eta}\right)^{(1)}_{{\bm k}_1}ik_{2i}f_{{\bm k}_2}^{(1)} &= q\left(\frac{\partial F}{\partial q}\right)^{(0)} \int\frac{{\rm d}^3 k_1 {\rm d}^3 k_2}{(2\pi)^3}\delta({\bm k}_1+{\bm k}_2-{\bm k}) D_1({\bm k}, {\bm k}_1, {\bm k}_2, {\bm \gamma}, \eta) \ \label{eq:A_component1},\\
    \left(\frac{{\rm d}q}{{\rm d}\eta}\right)^{(1)}_{{\bm k}_1}\frac{\partial f_{{\bm k}_2}^{(1)}}{\partial q} &= \left(q\left(\frac{\partial F}{\partial q}\right)^{(0)} + q^2\left(\frac{\partial^2 F}{\partial q^2}\right)^{(0)}\right) \int\frac{{\rm d}^3 k_1 {\rm d}^3 k_2}{(2\pi)^3}\delta({\bm k}_1+{\bm k}_2-{\bm k}) D_2({\bm k}, {\bm k}_1, {\bm k}_2, {\bm \gamma}, \eta) \ \label{eq:A_component2},\\
    \left(\frac{{\rm d}\gamma^i}{{\rm d}\eta}\right)^{(1)}_{{\bm k}_1}\frac{\partial f_{{\bm k}_2}^{(1)}}{\partial \gamma^i} &= q\left(\frac{\partial F}{\partial q}\right)^{(0)} \int\frac{{\rm d}^3 k_1 {\rm d}^3 k_2}{(2\pi)^3}\delta({\bm k}_1+{\bm k}_2-{\bm k}) D_3({\bm k}, {\bm k}_1, {\bm k}_2, {\bm \gamma}, \eta) \ \label{eq:A_component3},\\
    \left(\frac{{\rm d}q}{{\rm d}\eta}\right)^{(2)}_{{\bm k}} \left(\frac{\partial F}{\partial q}\right)^{(0)} &= q\left(\frac{\partial F}{\partial q}\right)^{(0)} \left[\int\frac{{\rm d}^3 k_1 {\rm d}^3 k_2}{(2\pi)^3}\delta({\bm k}_1+{\bm k}_2-{\bm k})D_{4{\rm \uppercase\expandafter{\romannumeral1}}}({\bm k}, {\bm k}_1, {\bm k}_2, {\bm \gamma}, \eta) + D_{4{\rm \uppercase\expandafter{\romannumeral2}}}({\bm k}, {\bm \gamma}, \eta)\right] \label{eq:A_component4}\ ,
\end{align}
where
\begin{equation}
    \begin{split}
        D_1({\bm k}, {\bm k}_1, {\bm k}_2, {\bm \gamma}, \eta') =& \left(\phi_{{\bm k}_1}^{(1)}(\eta')+\psi_{{\bm k}_1}^{(1)}(\eta')-\frac{1}{2}\gamma^r\gamma^s\epsilon_{{\bm k}_1,rs}^\lambda h_{{\bm k}_1,\lambda}^{(1)}(\eta')\right)ik_{2m}\gamma^m \times \\
        &\left[\phi_{{\bm k}_2}^{(1)}(\eta')+\int_{\eta_\mathrm{dec}}^{\eta'}{\rm d}{\eta''}\left(\frac{1}{2}\gamma^i\gamma^j\epsilon_{{\bm k}_2,ij}^{\lambda'} h_{{\bm k}_2,{\lambda'}}^{(1)'}(\eta'')-\phi_{{\bm k}_2}^{(1)'}(\eta'')-\psi_{{\bm k}_2}^{(1)'}(\eta'')\right)e^{-i\gamma^l k_{2l}(\eta'-\eta'')}\right]\ ,
    \end{split}
\end{equation}
\begin{equation}
    \begin{split}
        D_2({\bm k}, {\bm k}_1, {\bm k}_2, {\bm \gamma}, \eta') =& \left(-i\gamma^r k_{1r}\phi_{{\bm k}_1}^{(1)}(\eta') + \psi_{{\bm k}_1}^{(1)'}(\eta') -\frac{1}{2}\gamma^r\gamma^s\epsilon_{{\bm k}_1,rs}^\lambda h_{{\bm k}_1,\lambda}^{(1)'}(\eta')\right)\times\\
        &\left[\phi_{{\bm k}_2}^{(1)}(\eta')+\int_{\eta_\mathrm{dec}}^{\eta'}{\rm d}{\eta''}\left(\frac{1}{2}\gamma^i\gamma^j\epsilon_{{\bm k}_2,ij}^{\lambda'} h_{{\bm k}_2,{\lambda'}}^{(1)'}(\eta'')-\phi_{{\bm k}_2}^{(1)'}(\eta'')-\psi_{{\bm k}_2}^{(1)'}(\eta'')\right)e^{-i\gamma^l k_{2l}(\eta'-\eta'')}\right]\ ,
    \end{split}
\end{equation}
\begin{equation}
    \begin{split}
        D_3({\bm k}, {\bm k}_1, {\bm k}_2, {\bm \gamma}, \eta') =& \bigg[-ik_{1}^i(\phi^{(1)}_{{\bm k}_1}(\eta') + \psi^{(1)}_{{\bm k}_1}(\eta')) + ik_{1j}\gamma^i\gamma^j(\phi^{(1)}_{{\bm k}_1}(\eta') + \psi^{(1)}_{{\bm k}_1}(\eta')) \\
        &- \delta^{ij}\gamma^l\epsilon_{{\bm k}_1,lj}^\lambda h_{{\bm k}_1,\lambda}^{(1)'}(\eta') - i\delta^{ij}k_{1m}\gamma^m\gamma^l \epsilon_{{\bm k}_1,lj}^\lambda h_{{\bm k}_1,\lambda}^{(1)}(\eta') + \frac{1}{2}ik^i_1\gamma^l\gamma^m\epsilon_{{\bm k}_1,lm}^\lambda h^{(1)}_{{\bm k}_1,\lambda}(\eta')\\
        & + \gamma^i\gamma^l\gamma^m \epsilon_{{\bm k}_1,lm}^\lambda h^{(1)'}_{{\bm k}_1,\lambda}(\eta') + \frac{1}{2}ik_{1j}\gamma^i\gamma^j\gamma^l\gamma^m\epsilon_{{\bm k}_1,lm}^\lambda h_{{\bm k}_1,\lambda}^{(1)}(\eta') \bigg]\times \\
        &\int_{\eta_\mathrm{dec}}^{\eta'}{\rm d}\eta''\bigg[\gamma^r\epsilon_{{\bm k}_2,ir}^{\lambda'}h^{(1)'}_{{\bm k}_2,\lambda'}(\eta'')-ik_{2i}(\eta'-\eta'')\big[\frac{1}{2}\gamma^r\gamma^s\epsilon_{{\bm k}_2,rs}^{\lambda'}h_{{\bm k}_2,\lambda'}^{(1)'}(\eta'')\\
        &-\phi_{{\bm k}_2}^{(1)'}(\eta'')-\psi_{{\bm k}_2}^{(1)'}(\eta'')\big]\bigg]e^{-i\gamma^n k_{2n}(\eta'-\eta'')} \ ,
    \end{split}
\end{equation}
\begin{equation}
    \begin{split}
        D_{4{\rm \uppercase\expandafter{\romannumeral1}}}({\bm k}, {\bm k}_1, {\bm k}_2, {\bm \gamma}, \eta') =& -ik_{2i}\gamma^i\psi_{{\bm k}_1}^{(1)}(\eta')\phi_{{\bm k}_2}^{(1)}(\eta') + \phi_{{\bm k}_1}^{(1)}(\eta')\phi_{{\bm k}_2}^{(1)'}(\eta') + 2ik_{i}\gamma^i\phi_{{\bm k}_1}^{(1)}(\eta')\phi_{{\bm k}_2}^{(1)}(\eta')\\
        &+ ik_{2i}\gamma^i\gamma^r\gamma^s\epsilon_{{\bm k}_1,rs}^\lambda h_{{\bm k}_1,\lambda}^{(1)}(\eta')\phi_{{\bm k}_2}^{(1)}(\eta') + 2\psi_{{\bm k}_1}^{(1)}(\eta')\psi_{{\bm k}_2}^{(1)'}(\eta')\\
        &- \gamma^r\gamma^s\epsilon_{{\bm k}_1,rs}^\lambda h_{{\bm k}_1,\lambda}^{(1)'}(\eta')\psi_{{\bm k}_2}^{(1)}(\eta') - \gamma^r\gamma^s\epsilon_{{\bm k}_1,rs}^\lambda h_{{\bm k}_1,\lambda}^{(1)}(\eta')\psi_{{\bm k}_2}^{(1)'}(\eta')\\
        &+\frac{1}{2}\gamma^i\gamma^j\gamma^r\gamma^s\epsilon_{{\bm k}_1,ij}^\lambda\epsilon_{{\bm k}_2,ij}^{\lambda'}h_{{\bm k}_1,\lambda}^{(1)}(\eta')h_{{\bm k}_2,\lambda}^{(1)'}(\eta') \ ,
    \end{split}
\end{equation}
\begin{equation}
    \begin{split}
        D_{4{\rm \uppercase\expandafter{\romannumeral2}}}({\bm k}, {\bm k}_1, {\bm k}_2, {\bm \gamma}, \eta') =& \frac{1}{2}\psi_{{\bm k}}^{(2)'}(\eta') - \frac{1}{4}\gamma^i\gamma^j\epsilon_{{\bm k},ij}^\lambda h_{{\bm k},\lambda}^{(2)'}(\eta') - ik_i\gamma^i\phi_{{\bm k}}^{(2)}(\eta') + \frac{1}{2}ie_ik_j\gamma^i\gamma^jV^{(2)}_{\bm k}(\eta')\\
        &- \gamma^i e_i\mathcal{H}V_{\bm k}^{(2)}(\eta') \ .
    \end{split}
\end{equation}
Using eqs.~(\ref{eq:definition_of_A}) and (\ref{eq:A_component1})-(\ref{eq:A_component4}), we obtain 
\begin{equation}
    \begin{split}
        A(q,{\bm k},{\bm \gamma},\eta) =& -q\left(\frac{\partial F}{\partial q}\right)^{(0)}\left[\int\frac{{\rm d}k_1{\rm d}k_2}{(2\pi)^3}\delta({\bm k}_1+{\bm k}_2-{\bm k})(D_1+D_2+D_3+D_{4{\rm \uppercase\expandafter{\romannumeral1}}})+D_{4{\rm \uppercase\expandafter{\romannumeral2}}}\right]\\
        &- q^2\left(\frac{\partial^2 F}{\partial q^2}\right)^{(0)} \int\frac{{\rm d}k_1{\rm d}k_2}{(2\pi)^3}\delta({\bm k}_1+{\bm k}_2-{\bm k})D_2 \ .
    \end{split}
\end{equation}

Define
\begin{align}
    \tilde{D}_{n,\lambda}({\bm k}, ..., \eta) &\equiv \int_{\eta_\mathrm{dec}}^{\eta} {\rm d}\eta' \left[\frac{1}{4\pi}\int {\rm d}\Omega_q \gamma^i \gamma^j \epsilon_{{\bm k},\lambda,ij} D_{n}({\bm k}, ..., {\bm \gamma}, \eta') e^{-i k_r\gamma^r(\eta-\eta')}\right]\ ,\label{eq:def_of_Dtilde}
\end{align}
where $D_n$ denotes the functions $D_1$, $D_2$, $D_3$, $D_{4{\rm \uppercase\expandafter{\romannumeral1}}}$ and $D_{4{\rm \uppercase\expandafter{\romannumeral2}}}$. Following eqs.~(\ref{eq:Pi_ij_n}) and (\ref{eq:2o_f}), we obtain
\begin{equation}
    \begin{split}
        \sigma_{{\bm k},\lambda} =& 2\epsilon_{{\bm k}, \lambda}^{ij} a^{-4} \int \frac{{\rm d}^3 q}{(2\pi)^3} q\gamma_i\gamma_j \int_{\eta_\mathrm{dec}}^\eta {\rm d}\eta' A(q,{\bm k},{\bm \gamma},\eta')e^{-i\gamma^j k_j(\eta-\eta')}\\
        =&-2a^{-4} \int \frac{{\rm d}^3 q}{(2\pi)^3} \bigg[q^2\left(\frac{\partial F}{\partial q}\right)^{(0)}\left(\int\frac{{\rm d}k_1{\rm d}k_2}{(2\pi)^3}\delta({\bm k}_1+{\bm k}_2-{\bm k})(\tilde{D}_1+\tilde{D}_2+\tilde{D}_3+\tilde{D}_{4{\rm \uppercase\expandafter{\romannumeral1}}})+\tilde{D}_{4{\rm \uppercase\expandafter{\romannumeral2}}}\right)\\
        &+q^3\left(\frac{\partial^2 F}{\partial q^2}\right)^{(0)} \int\frac{{\rm d}k_1{\rm d}k_2}{(2\pi)^3}\delta({\bm k}_1+{\bm k}_2-{\bm k})\tilde{D}_2 \bigg]\\
        =&8\rho_\nu^{(0)} \bigg[\int\frac{{\rm d}^3 k_1 {\rm d}^3 k_2}{(2\pi)^3}\delta({\bm k}_1+{\bm k}_2-{\bm k})(\tilde{D}_{1,\lambda}-4\tilde{D}_{2,\lambda}+\tilde{D}_{3,\lambda}+\tilde{D}_{4{\rm \uppercase\expandafter{\romannumeral1}},\lambda})+\tilde{D}_{4{\rm \uppercase\expandafter{\romannumeral2}},\lambda}\bigg]\ .
    \end{split}
\end{equation}

\section{Bessel Function and the Angular Integration}\label{sec:bessel}
Following Rayleigh's formulas, one can write the spherical Bessel functions $j_n(x), n=0,1,2...$ as
\begin{equation}
    j_n(x)=(-x)^n\left(\frac{1}{x}\frac{\rm d}{{\rm d}x}\right)^n \frac{\sin x}{x} \ .
\end{equation}
Then we obtain the recurrence formula
\begin{equation}\label{eq:j_to_j}
    \frac{1}{x}\frac{\rm d}{{\rm d}x}\left(\frac{j_n(x)}{(-x)^n}\right) = \frac{j_{n+1}(x)}{(-x)^{n+1}} \ .
\end{equation}
Define the integrals
\begin{equation}
    I^{i_1i_2\dots i_N}_N ({\bm x})\equiv \frac{1}{4\pi} \int {\rm d}\Omega_q \gamma^{i_1} \gamma^{i_2} \dots \gamma^{i_N} e^{-i\gamma^j x_j} \ .
\end{equation}
We find that
\begin{equation}\label{eq:I_to_I}
    I_{N+1}^{i_1i_2\dots i_{N+1}} ({\bm x}) = i\frac{\partial}{\partial x_{i_{N+1}}}I_N^{i_1i_2\dots i_N} \ ,
\end{equation}
and 
\begin{equation}\label{eq:I_0}
    I_0 ({\bm x}) = \frac{1}{4\pi}\int {\rm d}\Omega_q e^{-i\gamma^j x_j} = \frac{\sin x}{x} = j_0(x) \ ,
\end{equation}
where $x=|{\bm x}|$. Using the recurrence formulas eqs.~(\ref{eq:j_to_j}), (\ref{eq:I_to_I}) and the initial condition eq.~(\ref{eq:I_0}), we get the general term formula of $I_N^{i_1 i_2 \dots i_N}({\bm x})$,
\begin{align}
    I_{2n+1}^{i_1 i_2 \dots i_{2n+1}}({\bm x}) &= i^{2n+1} \sum_{l=0}^{n} \delta^{(n-l)} x^{[2l+1]} \frac{j_{n+l+1}(x)}{(-x)^{n+l+1}} \label{eq:Int1} \ ,\\
    I_{2n}^{i_1 i_2 \dots i_{2n}} ({\bm x}) &= i^{2n} \sum_{l=0}^n \delta^{(n-l)} x^{[2l]} \frac{j_{n+l}(x)}{(-x)^{n+l}} \label{eq:Int2}\ ,
\end{align}
where $n=0,1,2\dots$, $\delta^{(n-l)}$ denotes $n-l$ Kronecker $\delta$'s with $2(n-l)$ distinct indexes, $x^{[2l]}$ denotes $2l$ x's with $2l$ distinct indexes. These $2n+1$ ($2n$) indexes are circulant symmetric. The explicit forms of eqs.~(\ref{eq:Int1}) and (\ref{eq:Int2}) for $n=1, 2$ have been shown in Ref.~\cite{book:75690}. From these equations, we obtain
\begin{align}
    I_2^{i_1 i_2}({\bm x})\epsilon_{{\bm k}, i_1 i_2}^\lambda &= - \frac{j_2(x)}{(-x)^2}x^{i_1}x^{i_2}\epsilon_{{\bm k}, i_1 i_2}^\lambda \ ,\\
    I_3^{i_1 i_2 i_3}({\bm x})\epsilon_{{\bm k},i_1 i_2}^\lambda & = -i\left(2\frac{j_2(x)}{(-x)^2}\delta^{i_1 i_3}x^{i_2} + \frac{j_3(x)}{(-x)^3}x^{i_1}x^{i_2}x^{i_3}\right)\epsilon_{{\bm k},i_1 i_2}^\lambda \ ,\\
    I_4^{i_1 i_2 i_3 i_4}({\bm x})\epsilon_{{\bm k}, i_1 i_2}^\lambda \epsilon_{{\bm k}_1, i_3 i_4}^{\lambda'} &= \left(2\frac{j_2(x)}{(-x)^2}\delta^{i_1 i_3}\delta^{i_2 i_4}+4\delta^{i_1 i_3}x^{i_2}x^{i_4}\frac{j_3(x)}{(-x)^3}+x^{i_1}x^{i_2}x^{i_3}x^{i_4}\frac{j_4(x)}{(-x)^4}\right)\epsilon_{{\bm k}, i_1 i_2}^\lambda \epsilon_{{\bm k}_1, i_3 i_4}^{\lambda'} \ ,\\
    &\cdots \ .
\end{align}

For $x\rightarrow \infty$, it should be noticed that 
\begin{align}
    \delta^{(n-l)} x^{[2l+1]} \frac{j_{n+l+1}(x)}{(-x)^{n+l+1}} &\ll \delta^{(n-l-1)} x^{[2l+3]} \frac{j_{n+l+2}(x)}{(-x)^{n+l+2}} \ ,\\
    \delta^{(n-l)} x^{[2l]} \frac{j_{n+l}(x)}{(-x)^{n+l}} &\ll \delta^{(n-l-1)} x^{[2l+2]} \frac{j_{n+l+1}(x)}{(-x)^{n+l+1}} \ .
\end{align}
Thus we have 
\begin{align}\label{eq:approximation}
    I_N^{i_1 i_2 \dots i_N}({\bm x}) &\simeq i^N x^{[N]} \frac{j_N(x)}{(-x)^N}
\end{align}
for $x\rightarrow\infty$.

\section{The Expressions of $f_i$, $\mathcal{P}_i$ and $\mathcal{Q}_i$}\label{sec:P_and_Q}

In the radiation dominated epoch, the explicit expression of $f_i$ are as follows,
\begin{equation}
    \begin{split}
        f_1 =& -2T_\phi(ux)T_\phi(vx) - 2T_\psi(ux)T_\phi(vx) + T_\psi(ux)T_\psi(vx) - 2vxT_\phi(ux)\frac{{\rm d}T_\psi(vx)}{{\rm d}(vx)}\\
        &- uvx^2\frac{{\rm d}T_\psi(ux)}{{\rm d}(ux)}\frac{{\rm d}T_\psi(vx)}{{\rm d}(vx)} \ ,
    \end{split}
\end{equation}

\begin{equation}
    \begin{split}
        f_{2{\rm \uppercase\expandafter{\romannumeral1}}}=&u^2\frac{{\rm d}^2\chi(ux)}{{\rm d}(ux)^2}T_\phi(vx) + \frac{2}{x}u\frac{{\rm d}\chi(ux)}{{\rm d}(ux)}T_\phi(vx) - \frac{2}{x}v\chi(ux)\frac{{\rm d}T_\phi(vx)}{{\rm d}(vx)}\\
        & -\frac{8}{x}v\chi(ux)\frac{{\rm d}T_\psi(vx)}{{\rm d}(vx)} - 3v^2 \chi(ux) \frac{{\rm d}^2 T_\psi(vx)}{{\rm d}(vx)^2} - \left(\frac{1}{4}u^2+\frac{23}{12}v^2+1\right)\chi(ux)T_\psi(vx) \\
        & + \frac{1}{4}(u^2+5v^2-1)\chi(ux)T_\phi(vx) \ ,
    \end{split}
\end{equation}

\begin{equation}\label{eq:f22}
    f_{2{\rm \uppercase\expandafter{\romannumeral2}}} = \chi(ux)(-T_\psi(vx)+T_\phi(vx)) \ ,
\end{equation}

\begin{equation}
    f_{3{\rm \uppercase\expandafter{\romannumeral1}}} = \frac{1}{2}uv\frac{{\rm d}\chi(ux)}{{\rm d}(ux)}\frac{{\rm d}\chi(vx)}{{\rm d}(vx)} + \frac{1}{4}(1-u^2-v^2)\chi(ux)\chi(vx) \ ,
\end{equation}

\begin{equation}\label{eq:f32}
    f_{3{\rm \uppercase\expandafter{\romannumeral2}}} = \frac{1}{2}\chi(ux)\chi(vx) \ ,
\end{equation}

\begin{equation}
    f_{3{\rm \uppercase\expandafter{\romannumeral3}}} = \chi(ux)\chi(vx) \ ,
\end{equation}

\begin{equation}
    f_{3{\rm \uppercase\expandafter{\romannumeral4}}} = -\frac{1}{4}\chi(ux)\chi(vx) \ ,
\end{equation}

\begin{equation}\label{eq:f35}
    f_{3{\rm \uppercase\expandafter{\romannumeral5}}} = -\frac{1}{2}\chi(ux)\chi(vx) \ .
\end{equation}

In the radiation dominated epoch, define $f_\nu$ as the fraction of the total energy density in neutrinos. The explicit expression of $\mathcal{P}_i$ are as follows,
\begin{equation}
    \begin{split}
        \mathcal{P}_1 =&\frac{12f_\nu}{x^2}\int_{x_\mathrm{dec}}^x {\rm d}x'\int_{x_\mathrm{dec}}^{x'}{\rm d}x''\bigg(2\frac{j_2(\tilde{x}_1)}{(-\tilde{x}_1)^2}(x'-x'') + \frac{1}{2}\frac{j_3(\tilde{x}_1)}{(-\tilde{x}_1)^3}(x'-x'')^2((1+v^2-u^2)x+(u^2+v^2-1)x'-2v^2 x'')\bigg)\\
        &\times v(T_\phi(ux')+T_\psi(ux'))\left(\frac{{\rm d}T_\phi(vx'')}{{\rm d}(vx'')}+\frac{{\rm d}T_\psi(vx'')}{{\rm d}(vx'')}\right) \\ 
        &-\frac{48f_\nu}{x^2}\int_{x_\mathrm{dec}}^x {\rm d}x'\int_{x_\mathrm{dec}}^{x'}{\rm d}x'' \bigg[\bigg(2\frac{j_2(\tilde{x}_1)}{(-\tilde{x}_1)^2}(x'-x'')-\frac{1}{2}\frac{j_3(\tilde{x}_1)}{(-\tilde{x}_1)^3}(x'-x'')^2((1+u^2-v^2)x-2u^2 x'+(u^2+v^2-1) x'')\bigg)\\
        &\times v T_\phi(ux') + \frac{j_2(\tilde{x}_1)}{(-\tilde{x}_1)^2}(x'-x'')^2 uv \frac{{\rm d}T_\psi(ux')}{{\rm d}(ux')} \bigg] \left(\frac{{\rm d}T_\phi(vx'')}{{\rm d}(vx'')}+\frac{{\rm d}T_\psi(vx'')}{{\rm d}(vx'')}\right) \\
        &+ \frac{12f_\nu}{x^2}\int_{x_\mathrm{dec}}^x {\rm d}x'\int_{x_\mathrm{dec}}^{x'}{\rm d}x'' \frac{1}{2}\frac{j_2(\tilde{x_1})}{(-\tilde{x_1})^2}(x'-x'')^3(1-u^2-v^2)v (T_\phi(ux')+T_\psi(ux'))\left(\frac{{\rm d}T_\phi(vx'')}{{\rm d}(vx'')}+\frac{{\rm d}T_\psi(vx'')}{{\rm d}(vx'')}\right)\\
        &- \frac{12f_\nu}{x^2}\int_{x_\mathrm{dec}}^x {\rm d}x'\int_{x_\mathrm{dec}}^{x'}{\rm d}x'' \bigg(2\frac{j_2(\tilde{x}_1)}{(-\tilde{x}_1)^2} + \frac{j_3(\tilde{x}_1)}{(-\tilde{x}_1)^3}(x'-x'')(2(-u^2+v^2)x+(3u^2+v^2-1)x'+(1-3v^2-u^2)x'')\\
        &+\frac{1}{2}\frac{j_3(\tilde{x}_1)}{(-\tilde{x}_1)^3}(u^2+v^2-1)(x'-x'')^2\\
        &-\frac{1}{4}\frac{j_4(\tilde{x}_1)}{(-\tilde{x}_1)^4}(x'-x'')^2((1+v^2-u^2)x + (u^2+v^2-1)x' - 2v^2 x'') ((u^2+1-v^2)x - 2u^2x' + (u^2+v^2-1)x'' ) \bigg) \\
        &\times v(x'-x'') (T_\phi(ux')+T_\psi(ux'))\left(\frac{{\rm d}T_\phi(vx'')}{{\rm d}(vx'')}+\frac{{\rm d}T_\psi(vx'')}{{\rm d}(vx'')}\right)\ ,
    \end{split}
\end{equation}

\begin{equation}\label{eq:P21}
    \begin{split}
        \mathcal{P}_{2{\rm \uppercase\expandafter{\romannumeral1}}} =& \frac{12f_\nu}{x^2}\int_{x_\mathrm{dec}}^x {\rm d}x' \frac{1}{2}\frac{j_3(x-x')}{(-(x-x'))^3}(v^2+1-u^2)(x-x')\chi(ux')T_\phi(vx') \\
        &- \frac{12f_\nu}{x^2} \int_{x_\mathrm{dec}}^x {\rm d}x' \int_{x_\mathrm{dec}}^{x'} {\rm d}x'' \frac{1}{2}\frac{j_3(\tilde{x}_1)}{(-\tilde{x}_1)^3}((1+v^2-u^2)x+(u^2+v^2-1)x'-2v^2x'')\chi(ux')v\left(\frac{{\rm d}T_\phi(vx'')}{{\rm d}(vx'')}+\frac{{\rm d}T_\psi(vx'')}{{\rm d}(vx'')}\right) \\
        &- \frac{12f_\nu}{x^2} \int_{x_\mathrm{dec}}^x {\rm d}x' \int_{x_\mathrm{dec}}^{x'} {\rm d}x'' \frac{1}{2}\frac{j_3(\tilde{x}_2)}{(-\tilde{x}_2)^3}((1+u^2-v^2)x+(u^2+v^2-1)x'-2u^2x'')\frac{{\rm d}\chi(ux'')}{{\rm d}(ux'')}u\left(T_\phi(vx')+T_\psi(vx')\right) \\
        &+ \frac{24f_\nu}{x^2} \int_{x_\mathrm{dec}}^x {\rm d}x' \frac{j_2(x-x')}{(-(x-x'))^2}u\frac{{\rm d}\chi(ux')}{{\rm d}(ux')}T_\phi(vx') \\
        &- \frac{24f_\nu}{x^2} \int_{x_\mathrm{dec}}^x {\rm d}x' \int_{x_\mathrm{dec}}^{x'} {\rm d}x'' \frac{j_2(\tilde{x}_2)}{(-\tilde{x}_2)^2}uv\frac{{\rm d}\chi(ux'')}{{\rm d}(ux'')}\frac{{\rm d}T_\phi(vx')}{{\rm d}(vx')} \\
        &- \frac{24f_\nu}{x^2} \int_{x_\mathrm{dec}}^x {\rm d}x' \int_{x_\mathrm{dec}}^{x'} {\rm d}x'' \frac{j_2(\tilde{x}_1)}{(-\tilde{x}_1)^2}uv\frac{{\rm d}\chi(ux')}{{\rm d}(ux')}\left(\frac{{\rm d}T_\phi(vx'')}{{\rm d}(vx'')}+\frac{{\rm d}T_\psi(vx'')}{{\rm d}(vx'')}\right) \\
        &+ \frac{12f_\nu}{x^2} \int_{x_\mathrm{dec}}^x {\rm d}x' \int_{x_\mathrm{dec}}^{x'} {\rm d}x'' \frac{1}{2}\frac{j_2(\tilde{x}_2)}{(-\tilde{x}_2)^2}(u^2+v^2-1)(x'-x'')u\frac{{\rm d}\chi(ux'')}{{\rm d}(ux'')}\left(T_\phi(vx')+T_\psi(vx')\right) \\
        &- \frac{12f_\nu}{x^2} \int_{x_\mathrm{dec}}^x {\rm d}x' 2 \frac{j_2(\tilde{x}_2)}{(-\tilde{x}_2)^2} \left( u \frac{{\rm d}\chi(ux')}{{\rm d}ux')} T_\psi(vx') + v \chi(ux')\frac{{\rm d}T_\psi(vx')}{{\rm d}(vx')} \right) \ ,
    \end{split}
\end{equation}

\begin{equation}
    \begin{split}
        \mathcal{P}_{2{\rm \uppercase\expandafter{\romannumeral2}}} =& - \frac{12f_\nu}{x^2} \int_{x_\mathrm{dec}}^x {\rm d}x' 2 \frac{j_3(\tilde{x-x'}_2)}{(-(x-x'))^3}(x-x')\chi(ux')T_\phi(vx')\\
        &- \frac{12f_\nu}{x^2} \int_{x_\mathrm{dec}}^x {\rm d}x' \int_{x_\mathrm{dec}}^{x'} {\rm d}x'' 2\frac{j_3(\tilde{x}_1)}{(-\tilde{x}_1)^3}(x+x'-2x'')\chi(ux')v\left(\frac{{\rm d}T_\phi(vx'')}{{\rm d}(vx'')}+\frac{{\rm d}T_\psi(vx'')}{{\rm d}(vx'')}\right) \\
        &- \frac{12f_\nu}{x^2} \int_{x_\mathrm{dec}}^x {\rm d}x' \int_{x_\mathrm{dec}}^{x'} {\rm d}x'' 2\frac{j_3(\tilde{x}_2)}{(-\tilde{x}_2)^3}(x-x')\frac{{\rm d}\chi(ux'')}{{\rm d}(ux'')}u\left(T_\phi(vx')+T_\psi(vx')\right) \\
        & - \frac{48f_\nu}{x^2} \int_{x_\mathrm{dec}}^x {\rm d}x' \int_{x_\mathrm{dec}}^{x'} {\rm d}x'' 2\frac{j_3(\tilde{x}_2)}{(-\tilde{x}_2)^3}(x-x')(x'-x'')uv\frac{{\rm d}\chi(ux'')}{{\rm d}(ux'')}\frac{{\rm d}T_\phi(vx')}{{\rm d}(vx')}\\
        & - \frac{48f_\nu}{x^2} \int_{x_\mathrm{dec}}^x {\rm d}x' \int_{x_\mathrm{dec}}^{x'} {\rm d}x'' 2\frac{j_3(\tilde{x}_1)}{(-\tilde{x}_1)^3}(x-x')(x'-x'')uv\frac{{\rm d}\chi(ux')}{{\rm d}(ux')}\left(\frac{{\rm d}T_\phi(vx'')}{{\rm d}(vx'')}+\frac{{\rm d}T_\psi(vx'')}{{\rm d}(vx'')}\right) \\
        &+ \frac{12f_\nu}{x^2} \int_{x_\mathrm{dec}}^x {\rm d}x' \int_{x_\mathrm{dec}}^{x'} {\rm d}x'' 2\frac{j_2(\tilde{x}_2)}{(-\tilde{x}_2)^2}(x'-x'')u\frac{{\rm d}\chi(ux'')}{{\rm d}(ux'')}\left(T_\phi(vx')+T_\psi(vx')\right) \\
        &+ \frac{12f_\nu}{x^2} \int_{x_\mathrm{dec}}^x {\rm d}x' \int_{x_\mathrm{dec}}^{x'} {\rm d}x'' \frac{j_3(\tilde{x}_2)}{(-\tilde{x}_2)^3}(x'-x'')^3 (1-u^2-v^2) u\frac{{\rm d}\chi(ux'')}{{\rm d}(ux'')}\left(T_\phi(vx')+T_\psi(vx')\right) \\
        &- \frac{12f_\nu}{x^2} \int_{x_\mathrm{dec}}^x {\rm d}x' \int_{x_\mathrm{dec}}^{x'} {\rm d}x'' 2 \frac{j_2(\tilde{x}_1)}{(-\tilde{x}_1)^2}(x'-x'')^2 uv \frac{{\rm d}\chi(ux')}{{\rm d}(ux')}\left(\frac{{\rm d}T_\phi(vx'')}{{\rm d}(vx'')}+\frac{{\rm d}T_\psi(vx'')}{{\rm d}(vx'')}\right) \ ,
    \end{split}
\end{equation}

\begin{equation}
    \mathcal{P}_{3{\rm \uppercase\expandafter{\romannumeral1}}} = \mathcal{P}_{3{\rm \uppercase\expandafter{\romannumeral2}}} = \mathcal{P}_{3{\rm \uppercase\expandafter{\romannumeral4}}} = 0 \ ,
\end{equation}

\begin{equation}
    \begin{split}
        \mathcal{P}_{3{\rm \uppercase\expandafter{\romannumeral3}}} =& -\frac{12f_\nu}{x^2} \int_{x_\mathrm{dec}}^x {\rm d}x' \int_{x_\mathrm{dec}}^{x'} {\rm d}x'' 2 \frac{j_3(\tilde{x}_2)}{(-\tilde{x}_2)^2}(x'-x'')(x-x')uv\frac{{\rm d}\chi(ux)}{{\rm d}(ux)}\frac{{\rm d}\chi(vx)}{{\rm d}(vx)} \\
        & + \frac{12f_\nu}{x^2} \int_{x_\mathrm{dec}}^x {\rm d}x' \int_{x_\mathrm{dec}}^{x'} {\rm d}x'' 2 \frac{j_3(\tilde{x}_1)}{(-\tilde{x}_1)^2}(x'-x'')(x-x'')uv\frac{{\rm d}\chi(ux)}{{\rm d}(ux)}\frac{{\rm d}\chi(vx)}{{\rm d}(vx)} \ ,
    \end{split}
\end{equation}

\begin{equation}\label{eq:P35}
    \mathcal{P}_{3{\rm \uppercase\expandafter{\romannumeral5}}} = \frac{12f_\nu}{x^2} \int_{x_\mathrm{dec}}^x {\rm d}x' \int_{x_\mathrm{dec}}^{x'} {\rm d}x'' 2\frac{j_3(\tilde{x}_1)}{(-\tilde{x}_1)^2} (x'-x'')^2 uv \frac{{\rm d}\chi(ux)}{{\rm d}(ux)}\frac{{\rm d}\chi(vx)}{{\rm d}(vx)} \ .
\end{equation}

In the radiation dominated epoch, the explicit expression of $\mathcal{Q}_i$ are expressed as,
\begin{equation}
    \begin{split}
        \mathcal{Q}_1 =&-\frac{48f_\nu}{x^2} \int_{x_\mathrm{dec}}^x {\rm d}x' \frac{1}{u} j_2(u(x-x'))\left(\frac{{\rm d}T_\phi(ux')}{{\rm d}(ux')}+\frac{{\rm d}T_\psi(ux')}{{\rm d}(ux')}\right)T_\psi(vx) \ ,
    \end{split}
\end{equation}

\begin{equation}
    \begin{split}
        \mathcal{Q}_{2{\rm \uppercase\expandafter{\romannumeral1}}} =&\frac{24f_\nu}{x^2} \int_{x_\mathrm{dec}}^x {\rm d}x' \frac{j_2(u(x-x'))}{(-u(x-x'))^2}u \frac{{\rm d}\chi(ux')}{{\rm d}(ux')}T_\psi(vx) \ ,
    \end{split}
\end{equation}

\begin{equation}
    \mathcal{Q}_{2{\rm \uppercase\expandafter{\romannumeral2}}} = \mathcal{Q}_{3j} = 0\ , \quad (j={\rm \uppercase\expandafter{\romannumeral1}}, {\rm \uppercase\expandafter{\romannumeral2}}, {\rm \uppercase\expandafter{\romannumeral3}}, {\rm \uppercase\expandafter{\romannumeral4}}, {\rm \uppercase\expandafter{\romannumeral5}})\ ,
\end{equation}
where ${\bm x}={\bm k}\eta$, ${\bm x}'={\bm k}\eta'$, ${\bm x}''={\bm k}\eta''$, $\tilde{{\bm x}}_1 = {\bm x} - u{\bm x}' - v{\bm x}''$, $\tilde{{\bm x}}_2 = {\bm x} - u{\bm x}'' - v{\bm x}'$, $\tilde{x}_1=|\tilde{{\bm x}}_1|$, $\tilde{x}_2=|\tilde{{\bm x}}_2|$. It should be clarified that during the calculations of $\mathcal{P}_{2l}$, $\mathcal{P}_{3m}$
($l={\rm \uppercase\expandafter{\romannumeral1}}, {\rm \uppercase\expandafter{\romannumeral2}}$, $m={\rm \uppercase\expandafter{\romannumeral1}}\sim {\rm \uppercase\expandafter{\romannumeral5}}$) shown above, the integrals $I_N^{i_1..i_N}$ have used the approximation for $N \geq 5$ shown in eq.~(\ref{eq:approximation}). Thus, eqs.~(\ref{eq:P21})-(\ref{eq:P35}) only established as $x\rightarrow \infty$.

\bibliography{biblio}

\begin{thebibliography}{10}

\bibitem{LIGOScientific:2016aoc}
B.~P. Abbott et~al.
\newblock {Observation of Gravitational Waves from a Binary Black Hole Merger}.
\newblock {\em Phys. Rev. Lett.}, 116(6):061102, 2016.

\bibitem{LIGOScientific:2016sjg}
B.~P. Abbott et~al.
\newblock {GW151226: Observation of Gravitational Waves from a 22-Solar-Mass
  Binary Black Hole Coalescence}.
\newblock {\em Phys. Rev. Lett.}, 116(24):241103, 2016.

\bibitem{Domenech:2021ztg}
Guillem Dom\`enech.
\newblock {Scalar Induced Gravitational Waves Review}.
\newblock {\em Universe}, 7(11):398, 2021.

\bibitem{Planck:2018vyg}
N.~Aghanim et~al.
\newblock {Planck 2018 results. VI. Cosmological parameters}.
\newblock {\em Astron. Astrophys.}, 641:A6, 2020.
\newblock [Erratum: Astron.Astrophys. 652, C4 (2021)].

\bibitem{Planck:2018jri}
Y.~Akrami et~al.
\newblock {Planck 2018 results. X. Constraints on inflation}.
\newblock {\em Astron. Astrophys.}, 641:A10, 2020.

\bibitem{Cai:2019bmk}
Rong-Gen Cai, Zong-Kuan Guo, Jing Liu, Lang Liu, and Xing-Yu Yang.
\newblock {Primordial black holes and gravitational waves from parametric
  amplification of curvature perturbations}.
\newblock {\em JCAP}, 06:013, 2020.

\bibitem{Saito:2008jc}
Ryo Saito and Jun'ichi Yokoyama.
\newblock {Gravitational wave background as a probe of the primordial black
  hole abundance}.
\newblock {\em Phys. Rev. Lett.}, 102:161101, 2009.
\newblock [Erratum: Phys.Rev.Lett. 107, 069901 (2011)].

\bibitem{Wang:2016ana}
Sai Wang, Yi-Fan Wang, Qing-Guo Huang, and Tjonnie G.~F. Li.
\newblock {Constraints on the Primordial Black Hole Abundance from the First
  Advanced LIGO Observation Run Using the Stochastic Gravitational-Wave
  Background}.
\newblock {\em Phys. Rev. Lett.}, 120(19):191102, 2018.

\bibitem{Wang:2019kaf}
Sai Wang, Takahiro Terada, and Kazunori Kohri.
\newblock {Prospective constraints on the primordial black hole abundance from
  the stochastic gravitational-wave backgrounds produced by coalescing events
  and curvature perturbations}.
\newblock {\em Phys. Rev. D}, 99(10):103531, 2019.
\newblock [Erratum: Phys.Rev.D 101, 069901 (2020)].

\bibitem{Yuan:2019udt}
Chen Yuan, Zu-Cheng Chen, and Qing-Guo Huang.
\newblock {Probing primordial\textendash{}black-hole dark matter with scalar
  induced gravitational waves}.
\newblock {\em Phys. Rev. D}, 100(8):081301, 2019.

\bibitem{Garcia-Bellido:2016dkw}
Juan Garcia-Bellido, Marco Peloso, and Caner Unal.
\newblock {Gravitational waves at interferometer scales and primordial black
  holes in axion inflation}.
\newblock {\em JCAP}, 12:031, 2016.

\bibitem{Cai:2018dig}
Rong-gen Cai, Shi Pi, and Misao Sasaki.
\newblock {Gravitational Waves Induced by non-Gaussian Scalar Perturbations}.
\newblock {\em Phys. Rev. Lett.}, 122(20):201101, 2019.

\bibitem{Inomata:2019zqy}
Keisuke Inomata, Kazunori Kohri, Tomohiro Nakama, and Takahiro Terada.
\newblock {Gravitational Waves Induced by Scalar Perturbations during a Gradual
  Transition from an Early Matter Era to the Radiation Era}.
\newblock {\em JCAP}, 10:071, 2019.

\bibitem{Inomata:2019ivs}
Keisuke Inomata, Kazunori Kohri, Tomohiro Nakama, and Takahiro Terada.
\newblock {Enhancement of Gravitational Waves Induced by Scalar Perturbations
  due to a Sudden Transition from an Early Matter Era to the Radiation Era}.
\newblock {\em Phys. Rev. D}, 100(4):043532, 2019.

\bibitem{Assadullahi:2009nf}
Hooshyar Assadullahi and David Wands.
\newblock {Gravitational waves from an early matter era}.
\newblock {\em Phys. Rev. D}, 79:083511, 2009.

\bibitem{Papanikolaou:2020qtd}
Theodoros Papanikolaou, Vincent Vennin, and David Langlois.
\newblock {Gravitational waves from a universe filled with primordial black
  holes}.
\newblock {\em JCAP}, 03:053, 2021.

\bibitem{Domenech:2020ssp}
Guillem Dom\`enech, Chunshan Lin, and Misao Sasaki.
\newblock {Gravitational wave constraints on the primordial black hole
  dominated early universe}.
\newblock {\em JCAP}, 04:062, 2021.

\bibitem{Chang:2020mky}
Zhe Chang, Sai Wang, and Qing-Hua Zhu.
\newblock {On the Gauge Invariance of Scalar Induced Gravitational Waves: Gauge
  Fixings Considered}.
\newblock 10 2020.

\bibitem{Chang:2020iji}
Zhe Chang, Sai Wang, and Qing-Hua Zhu.
\newblock {Gauge Invariant Second Order Gravitational Waves}.
\newblock 9 2020.

\bibitem{Chang:2020tji}
Zhe Chang, Sai Wang, and Qing-Hua Zhu.
\newblock {Note on gauge invariance of second order cosmological
  perturbations}.
\newblock {\em Chinese Physics C}, 45(9):095101, 2021.

\bibitem{Hwang:2017oxa}
Jai-Chan Hwang, Donghui Jeong, and Hyerim Noh.
\newblock {Gauge dependence of gravitational waves generated from scalar
  perturbations}.
\newblock {\em Astrophys. J.}, 842(1):46, 2017.

\bibitem{Zhou:2021vcw}
Jing-Zhi Zhou, Xukun Zhang, Qing-Hua Zhu, and Zhe Chang.
\newblock {The Third Order Scalar Induced Gravitational Waves}.
\newblock 6 2021.

\bibitem{Chang:2022dhh}
Zhe Chang, Xukun Zhang, and Jing-Zhi Zhou.
\newblock {The cosmological vector modes from a monochromatic primordial power
  spectrum}.
\newblock 7 2022.

\bibitem{Weinberg:2003ur}
Steven Weinberg.
\newblock {Damping of tensor modes in cosmology}.
\newblock {\em Phys. Rev. D}, 69:023503, 2004.

\bibitem{Dent:2013asa}
James~B. Dent, Lawrence~M. Krauss, Subir Sabharwal, and Tanmay Vachaspati.
\newblock {Damping of Primordial Gravitational Waves from Generalized Sources}.
\newblock {\em Phys. Rev. D}, 88:084008, 2013.

\bibitem{Watanabe:2006qe}
Yuki Watanabe and Eiichiro Komatsu.
\newblock {Improved Calculation of the Primordial Gravitational Wave Spectrum
  in the Standard Model}.
\newblock {\em Phys. Rev. D}, 73:123515, 2006.

\bibitem{Bartolo:2010qu}
N.~Bartolo, S.~Matarrese, and A.~Riotto.
\newblock {Non-Gaussianity and the Cosmic Microwave Background Anisotropies}.
\newblock {\em Adv. Astron.}, 2010:157079, 2010.

\bibitem{Mangilli:2008bw}
Anna Mangilli, Nicola Bartolo, Sabino Matarrese, and Antonio Riotto.
\newblock {The impact of cosmic neutrinos on the gravitational-wave
  background}.
\newblock {\em Phys. Rev. D}, 78:083517, 2008.

\bibitem{Saga:2014jca}
Shohei Saga, Kiyotomo Ichiki, and Naoshi Sugiyama.
\newblock {Impact of anisotropic stress of free-streaming particles on
  gravitational waves induced by cosmological density perturbations}.
\newblock {\em Phys. Rev. D}, 91(2):024030, 2015.

\bibitem{Nakama:2016enz}
Tomohiro Nakama and Teruaki Suyama.
\newblock {Primordial black holes as a novel probe of primordial gravitational
  waves. II: Detailed analysis}.
\newblock {\em Phys. Rev. D}, 94(4):043507, 2016.

\bibitem{Gong:2019mui}
Jinn-Ouk Gong.
\newblock {Analytic Integral Solutions for Induced Gravitational Waves}.
\newblock {\em Astrophys. J.}, 925(1):102, 2022.

\bibitem{Bartolo:2004if}
N.~Bartolo, E.~Komatsu, Sabino Matarrese, and A.~Riotto.
\newblock {Non-Gaussianity from inflation: Theory and observations}.
\newblock {\em Phys. Rept.}, 402:103--266, 2004.

\bibitem{book:2362664}
Michele Maggiore.
\newblock {\em Gravitational Waves, Volume 2: Astrophysics and Cosmology},
  volume~2.
\newblock Oxford University Press, 2018.

\bibitem{Kasai:1985jg}
Masumi Kasai and Kenji Tomita.
\newblock {Gauge Invariant Perturbations in a Universe With a Collisionless Gas
  and a Fluid}.
\newblock {\em Phys. Rev. D}, 33:1576, 1986.

\bibitem{Stefanek:2012hj}
Ben~A. Stefanek and Wayne~W. Repko.
\newblock {Analytic description of the damping of gravitational waves by free
  streaming neutrinos}.
\newblock {\em Phys. Rev. D}, 88(8):083536, 2013.

\bibitem{Pitrou:2013hga}
Cyril Pitrou, Xavier Roy, and Obinna Umeh.
\newblock {xPand: An algorithm for perturbing homogeneous cosmologies}.
\newblock {\em Class. Quant. Grav.}, 30:165002, 2013.

\bibitem{Kohri:2018awv}
Kazunori Kohri and Takahiro Terada.
\newblock {Semianalytic calculation of gravitational wave spectrum nonlinearly
  induced from primordial curvature perturbations}.
\newblock {\em Phys. Rev. D}, 97(12):123532, 2018.

\bibitem{book:15453}
Scott Dodelson.
\newblock {\em Modern cosmology}.
\newblock Academic Press, 1 edition, 2003.

\bibitem{Yuan:2019wwo}
Chen Yuan, Zu-Cheng Chen, and Qing-Guo Huang.
\newblock {Log-dependent slope of scalar induced gravitational waves in the
  infrared regions}.
\newblock {\em Phys. Rev. D}, 101(4):043019, 2020.

\bibitem{Espinosa:2017sgp}
J.~R. Espinosa, D.~Racco, and A.~Riotto.
\newblock {Cosmological Signature of the Standard Model Higgs Vacuum
  Instability: Primordial Black Holes as Dark Matter}.
\newblock {\em Phys. Rev. Lett.}, 120(12):121301, 2018.

\bibitem{Kuroda:2015owv}
Kazuaki Kuroda, Wei-Tou Ni, and Wei-Ping Pan.
\newblock {Gravitational waves: Classification, Methods of detection,
  Sensitivities, and Sources}.
\newblock {\em Int. J. Mod. Phys. D}, 24(14):1530031, 2015.

\bibitem{book:75690}
Steven Weinberg.
\newblock {\em Cosmology}.
\newblock Oxford University Press, USA, oxford university press, usa edition,
  2008.

\end{thebibliography}

\end{document}